\documentclass[12pt]{article}

\usepackage{amssymb,amsmath,amsfonts,authblk,booktabs,blindtext, eurosym,geometry,ulem,graphicx,caption,color,setspace,sectsty,comment,footmisc,caption,pdflscape,subfigure,array,hyperref,float, multirow, pbox, rotating, adjustbox, caption, regexpatch, siunitx, setspace,tabularx}
\usepackage[english]{babel}
\usepackage[flushleft]{threeparttable}
\usepackage{array}
\usepackage{ragged2e}
\usepackage{multicol}
\usepackage{verbatim} 
\usepackage{graphicx}
\usepackage{adjustbox}

\definecolor{darkblue}{rgb}{0.0,0.0,0.5}

\newcolumntype{P}[1]{>{\RaggedRight\hspace{0pt}}p{#1}}

\hypersetup{
	colorlinks= true,
	linkcolor = darkblue,
	urlcolor  = darkblue,
	citecolor = darkblue,
	anchorcolor = blue
}
\usepackage{csquotes}
\usepackage[style = apa, sorting = nyt,
			natbib = true,
			doi = false,
			isbn = false]{biblatex}
\title{Stablecoins: Survivorship, Transactions Costs and Exchange Microstructure}
\author[1]{Bruce Mizrach \thanks{Correspondence: Department of Economics, Rutgers University, 75 Hamilton Street, New Brunswick, NJ 08901 USA. email:
mizrach@econ.rutgers.edu, (908) 913-0253 (voice) and (425) 795-9942 (fax). 
Declarations: Availability of data and materials: Data are publicly available from Kaggle: https://www.kaggle.com/bigquery/ethereum-blockchain;
Competing interests: Not applicable; Funding: Not applicable; Author's contributions: Single author;
Acknowledgements: I would like to thank Tony Tassell, seminar participants at FINRA and the ECB, and five anonymous referees for helpful comments.}
\affil[1]{\small Department of Economics, Rutgers University, New Brunswick, NJ USA}}
\date{Revised: February 2023}
\begin{document}
\begin{titlepage}
\maketitle
\begin{abstract}
\noindent Stable coins are not very stable. Cash collateralized coins are more stable, but the overall failure rate is similar to tokens that are not designed to be stable. USD Coin, Tether and Dai have the largest Ethereum market shares, and they have an average velocity nearly three times higher than M1.   Centralized and decentralized exchanges are the most active nodes and largest holders on the blockchain. Four of the top ten tokens have Herfindahl indices higher than the U.S. banking system. Median gas fees for Tether rose more than twelve times over the last two years, and nearly twenty times for USD Coin.  Transactions of under \$50,000 can generally be done more cheaply off-chain. 24 hour exchange turnover in Tether is nearly \$60 billion.  This is comparable to the daily volume at the NYSE and eight times the daily flow in money market mutual funds.  Narrow bid-ask spreads and depth have attracted HFT participation approaching 50\%.

\end{abstract}

\setcounter{page}{0}
\thispagestyle{empty}
\end{titlepage}


\pagebreak \newpage
\setstretch{1.05}
\section{Introduction}

The financial industry has continued to develop new forms of payment.  Credit and debit cards and automated clearinghouse (ACH) transactions were the key developments of the late 20th century.  Venmo and PayPal came along in the early 21st.

Credit and debit cards still dominate, making up 62\% of purchases.\footnote{Aaron Black, "Meet the New Payment Champions Same as the Old Ones, \textit{Wall Street Journal}, January 11, 2019.}  Transaction costs are still substantial.  Credit card transactions are nearly instantaneous, but the merchants pay fees of 1 to 3\%.\footnote{\url{https://paymentdepot.com/blog/average-credit-card-processing-fees/}}  Debit cards fees are closer to 1\%,\footnote{\url{https://www.cardfellow.com/blog/debit-card-transaction-fees/}} yet consumers can face stiff payments for overdrawing their account.  Venmo is very convenient, but it also charges 3\% for credit card transactions.\footnote{\url{https://venmo.com/about/fees}}  Paypal's recent growth has largely been driven by opening up to credit card companies.

There were 7.3 billion ACH transactions with a total value of \$18.4 trillion in the second quarter of 2021\footnote{\url{https://www.nacha.org/news/strong-growth-continues-ach-network-volume-climbs-nearly-10-second-quarter-2021}} with a median cost of \$0.29.\footnote{Association for Financial Professionals, \url{https://www.afponline.org/docs/default-source/default-document-library/pub/2015-payments-cost-benchmarking-report}}
ACH transactions are quite slow though, and can take up to five days. The \citet{G7Stable} notes that there are still many gaps in the payment system, particularly for cross-border transactions.  With respect to fees, speed and network size, there are substantial opportunities for new financial technologies.  

Stablecoins try to harness the distributed ledger technology of Bitcoin and other digital assets, while maintaining  price stability of fiat assets and other stores of value.  \citet{Ante} provide a broad overview of the empirical literature on  the  sector. Tokens are generally released as \textit{smart contracts}, software code that enables market participants to exchange the tokens on the blockchain.

Ethereum protocols have become the dominant smart contract framework, and this has led most of the leading stable coins to adopt the ERC-20 (Ethereum Request for Comments-20) standard, originally proposed by Fabian Vogelsteller in November 2015.\footnote{\url{https://ethereum.org/en/developers/docs/standards/tokens/erc-20/}.}  The standard makes tokens interoperable, and it requires that the total supply of tokens and the balance of any address be transparent to the network.  For this reason, I focus my analysis on the Ethereum Mainnet. 

Etherscan reports\footnote{\url{https://etherscan.io/tokens}  Retrieved on May 16, 2022.} 529, 212 token contracts, including 945 tokens based on the ERC-20 standard.

\autoref{tab:stable_marketcap} contains the top ten Ethereum network stablecoins, pegged to the U.S. dollar, ranked by market cap as of the end of the first quarter of 2022.

\begin{table}[H]
  \centering
  \begin{threeparttable}
  \caption{Top Ten Stablecoins by Market Capitalization}
    \begin{tabular}{llllr}
\hline
    Name  & Symbol & First Transaction & Collateral & \multicolumn{1}{l}{Market Cap \$bn.} \\
\hline \hline
    USD Coin & USDC  & 2018-09-10 & Asset backed & \$45.87 \\
    Tether USD & USDT  & 2017-11-28 & Asset backed & \$39.82 \\
    Binance USD & BUSD  & 2019-09-10 & Asset backed & \$17.45 \\
    Dai Stablecoin & DAI   & 2019-11-13 & Defi  & \$9.22 \\
    TrueUSD & TUSD  & 2019-01-04 & Asset backed & \$1.09 \\
    Paxos Standard & PAX   & 2018-09-10 & Asset backed & \$0.97 \\
    Terra USD & UST   & 2020-12-03 & Defi, algo stabilized & \$0.36 \\
    HUSD  & HUSD  & 2019-01-04 & Asset backed & \$0.28 \\
    Gemini dollar & GUSD  & 2018-09-09 & Asset backed & \$0.25 \\
    Synth sUSD & sUSD  & 2020-05-11 & Defi, algo stabilized & \$0.14 \\
\hline
    \end{tabular}%
  \label{tab:stable_marketcap}
     \begin{tablenotes}
      \small
      \item The market capitalization is based on the circulating supply on the Ethereum network at the end of 2022Q1.
    \end{tablenotes}
    \end{threeparttable}%
\end{table}%

Across blockchain networks,\footnote{Several tokens including Tether are issued on more than one network.  I am showing the supply on the Ethereum blockchain.} there are more than \$150 billion in stablecoins outstanding.\footnote{https://coinmarketcap.com/view/stablecoin/, retrieved on September 26, 2022.}  Based on total assets, the stablecoin sector would be among the thirty largest banks in the United States.\footnote{https://www.federalreserve.gov/releases/lbr/current/}

The size and growth of the stablecoin sector has attracted the attention of regulators.  The President's Working Group on Financial Markets,\footnote{\citet{WorkingGroup}} which included representatives from the Securities and Exchange Commission (SEC), the Federal Deposit Insurance Commission (FDIC), the Office of the Comptroller of the Currency (OCC), the Commodities Futures Trading Commission (CFTC), and the Federal Reserve, has proposed that stable coin issuers become insured depository institutions.

This manuscript assesses the stablecoin sector in the context of comparable financial institutions and instruments.  I first examine the feasibility of the Working Group's recommendation with respect to collateral.  I find many shortcomings with both quality and disclosure, even with the leading coins.

I establish that stablecoins with cash equivalents as collateral are more stable.  This finding is now reinforced by the collapse of Terra USD in May 2022 which I discuss in Section 2.

I address the stability concerns of the regulators by also looking at hazard functions for stablecoins on the Ethereum network.  \citet{Liao} emphasize the risk of runs even in fully collateralized stablecoins.  Since 2016, more than 60\% of the stablecoins that reached the Mainnet have failed.  Stablecoins have similar failure rates with expected lifetimes that are only four months longer than non-stablecoins.

With respect to transactions, I will compare the stablecoins to monetary instruments. I compute the growth of transfer volumes on the Ethereum network.  The top ten coins had nearly \$1.5 trillion in transfer volume in the first quarter of 2022.  The velocity of circulation for the leading coins is two to three times as large as the M1 velocity.

The blockchain literature has made a distinction between centralized exchanges like Coinbase and decentralized exchanges (DEX) like Uniswap. \citet{Fritsch} notes the two key aspects of DEX: they allow users to retain custody of their tokens; and they utilize automated
market makers. \citet{Mohan} models and contrasts a variety of the current DEX trading protocols. A far larger volume of coins are traded in the off-blockchain centralized exchanges.  Total volume in Tether alone exceeds the daily turnover in the FANG.\footnote{Facebook (now Meta), Amazon, Netflix and Google (now Alphabet)}

The Ethereum network is more transparent than the interbank or equity market networks.  Any participant in the network can see all of the transactions in and out of Ethereum stablecoins at the level of a blockchain hashtag.

The majority of stablecoin transfers are between venues for trading digital assets.  Eight of ten most active network addresses for Tether are trading venues including Binance, Coinbase, FTX and Uniswap. USD Coin is similar with Coinbase, FTX and Uniswap the three most active addresses. \citet{MakarovBitcoin} find a similar result for Bitcoin, noting that 75\% of transfers are for trading or speculative purposes.

 I compute Herfindahl indices using the address network for the top 50 holders.  Holdings of the stablecoins can be highly concentrated, with Herfindahl indices ranging from 0.0088 for Tether to 0.4084 for Binance USD.  A single Binance address holds more than 57\% of its' total supply. Based on merger guidelines for the Department of Justice,\footnote{\url{https://www.justice.gov/atr/herfindahl-hirschman-index}} three top stablecoins, Binance USD, Gemini, and Huobi would be considered ``highly concentrated.''
 
 Because they may eventually become an important part of the payments system, I analyze transaction costs for stablecoin transfers.  Aggregate fees in the first quarter of 2022 totaled over \$336 million U.S. dollars.  Median fees rose more than ten-fold for the leading stablecoins in the last two years.
 
 I compare these transaction costs on the Mainnet to costs on centralized exchanges. Incorporating bid-ask spreads and exchange fees, I show that transactions on chain are generally cheaper for transfers of larger than \$50,000.  Transfers on chain are much slower though, and high-frequency trading is almost entirely on centralized exchanges.
 
 I begin my discussion of the stablecoin sector in Section 2 with an analysis of the collateral holdings.  I then turn to survival analysis in Section 3. I examine transaction volumes on and off-chain in Section 4.  My network analysis of active addresses is in Section 5.  Section 6 examines on-chain transaction costs both cross-sectionally and over time.  Section 7 computes transaction cost estimates for off-chain transactions on centralized exchanges and compares them to the costs on-chain.  Section 8 looks at speed and the amount of high frequency trading activity in the centralized exchanges.  Section 9 concludes.

\section{Collateral: Stablecoins v. Money Market Funds}
\citet{StableLaw} note that there are three major types of stablecoins: (1) Fiat/commodity collateralized; (2) Crypto-collateralized; (3) Non-collateralized.  Each type differs on the collateral backing the tokens.  Among the top tokens I study, the coins are either fiat or crypto-collateralized.

Collateralized stable coin companies are expected to actually hold the assets against which their coin is pegged (e.g., US dollar or gold). They issue new units as they expand their underlying assets.\footnote{\url{https://academy.binance.com/en/glossary/stablecoin}}

The valuation of crypto-collateralized coins is maintained through over-collateralization and stability mechanisms. In the case of Dai, smart contracts called Collateralized Debt Positions bring Dai into circulation.  You can only retrieve your collateral by paying back the debt.\footnote{Why is Dai Stable?, \url{https://medium.com/icovo/why-is-dai-stable-9a9fa84feca7}}

Non-collateralized stable coins use algorithms to dynamically expand and contract the supply of tokens to maintain a predetermined peg.  There are a number of prominent failures:  SagaCoin (SAGA) which replicated the IMF SDR;\footnote{\url{https://www.sogur.com/}} Havven (HAV), now rebranded as Synthetix;\footnote{\url{https://blog.havven.io/}} and Basis,\footnote{\url{https://www.coindesk.com/basis-stablecoin-confirms-shutdown-blaming-regulatory-constraints}} call into question the viability of the mechanism.

\subsection{Tether}
Tether was created by iFinex, the same company that runs the Bitfinex exchange and was originally called ``RealCoin."  It was launched on the Bitcoin blockchain using the Omni Layer Protocol on October 6, 2014.\footnote{\citet{Blockreport}}  The token I study here is an ERC-20 token that was issued after a \$31 million hack in late 2017.\footnote{\url{https://tether.to/2017/11/}}  I first detect the token on the Mainnet on November 28, 2017. As of May 2022, 50.5\% of Tether is on the Tron network, 45.4\% is on Ether, 1.6\% on Solana, and 1.0\% is still on Ommi. An additional 0.9\% is on six smaller networks including Avalanche, Algorand, EOS, Liquid, SLP and Statemine.\footnote{These data were retrieved from \url{wallet.tether.to/transparency} on May 11, 2022.}

The company originally represented Tether as something like a currency board: holding ``one U.S. dollar (“USD”) in reserve `backing' the tether... Every tether is always backed 1-to-1, by traditional currency held in our reserves. So 1 USDT is always equivalent to 1 USD.”

In March 2019, Tether stepped back from this claim instead reporting that:\footnote{\url{https://investingnews.com/daily/tech-investing/blockchain-investing/tether-admits-not-fully-backed-us-dollar/}} “Every tether is always 100 percent backed by our reserves, which include traditional currency and cash equivalents and, from time to time, may include other assets and receivables from loans made by Tether to third parties, which may include affiliated entities..."

The New York Attorney (NYAG) pursued a fraud case\footnote{\citet{NYAGTether}} against both Bitfinex and Tether, documenting a series of questionable bank transfers.  The NYAG has banned trading with New York persons and entities and fined the company \$18.5 million. The Commodity Futures Trading Commission also fined iFinex an additional \$41 million for misrepresentation of their collateral.\footnote{\url{https://www.cftc.gov/PressRoom/PressReleases/8450-21}} \citet{GriffinTether} have shown that Tether's price was manipulated to support Bitcoin.

In accordance with the ruling by the NYAG, Tether released a breakdown of their reserves on March 31, 2021.  In their report report\footnote{\url{https://tether.to/en/transparency/\#reports}} from 2021Q4, Tether held 83.74\% of its assets in cash and cash equivalents, 36.7\% of which was held in commercial paper.

During the collapse of Terra USD, which I discuss below, Tether's peg also came under pressure.  Tether responded by reducing the outstanding supply, making their collateral more liquid, and disclosing more information to investors.  In their report for August 2022 by BDO, a global accounting firm, Tether has reduced commercial paper holdings to 15.9\% of their cash equivalents.  Cash equivalents have actually fallen to 79.62\% of total assets though.

\subsection{USD Coin}

USD Coin is a stable coin developed by the fintech company Circle and the digital asset exchange Coinbase.  Coinbase originally claimed that each USD Coin is backed by U.S. dollars in ``a bank account."\footnote{\url{https://www.coinbase.com/usdc}: ``Each USDC is backed by one US dollar, which is held in a bank account.  Using the Internet time machine, I can verify this claim was on their website until July 20, 2021.}  Following an an auditor's report\href{https://www.centre.io/hubfs/pdfs/attestation/Grant-Thorton_circle_usdc_reserves_07162021.pdf}  by Grant Thornton on July 16, 2021, Coinbase modified this claim to say that ``Each USDC is backed by one dollar or asset with equivalent fair value, which is held in accounts with US regulated financial institutions.''

On August 22, 2021, Circle \href{https://www.centre.io/blog/usdc-reserves-composition}{announced} that ``it will now hold the USDC reserve entirely in cash and short duration US Treasuries.'' Attestation reports since September 30, 2021 are consistent with this claim.

USD Coin, with 48\% of transaction volume, along with Tether's 32\%, dominate the stablecoin market.  

\subsection{Binance USD}

Binance USD (BUSD) is a USD-denominated stable coin approved by the New York State Department of Financial Services (NYDFS) launched in September 2019 in partnership with Paxos and Binance.

The NYDFS maintains a ``green list" of approved virtual currencies.\footnote{\url{https://www.dfs.ny.gov/apps_and_licensing/virtual_currency_businesses/virtual_currencies}}  At the moment, the list includes thirteen currencies including Bitcoin, Ethereum and Litecoin.  The two leading stable coins on the list are Binance USD and Paxos (both standard and gold are approved).

Paxos is the custodian and issuer of the BUSD.  Paxos began to provide CUSIP level detail on the holdings of both BUSD and USDP in June 2022.\footnote{\url{https://paxos.com/busd-transparency/}}. As of August 31, 2022, BUSD holds 17 Treasury bills, 32 U.S. Treasury repurchase agreements on notes and bonds, and bank deposits at a set of possible FDIC insured financial institutions.  They hold private insurance against deposits that exceed the FDIC limits.

\subsection{Dai}

\citet{BahachukDai} notes that Dai was conceptualized by a group called MakerDAO. It is built on a decentralized Ethereum technology called the Maker Protocol.  DAO stands for Decentralized Autonomous Organization. The DAO holds four types of collateral, Ethereum, Basic Attention Token (BAT), wrapped Bitcoin, and USD Coin. There are strong financial incentives to keep the currency over-collateralized.

\citet{KozhanDao} model the DAO stabilization mechanism both theoretically and empirically.  They find that while the Dai price covaries negatively with risky collateral, the introduction of USDC as collateral has led to an increase in peg stability.

\subsection{TrueUSD}

TrueUSD provides real-time reports on the state of their collateral through the fintech firm Armanino.\footnote{\url{https://www.armaninollp.com/software/trustexplorer/real-time-audit/}}  At the time I first tried to retrieve the report in March 2021,  the auditing firm supplied a ``ripcord" message, indicating that they had not been able to verify the collateral for more than 72 hours.  In private communication with Armanino, I received the following reply:  ``TrueUSD was sold to a 3rd-party and therefore the AT-C 205 reporting has ceased until the new owners have completed the onboarding process. All other TrueCurrencies are still owned by TrustToken and have live TrueCurrencies."\footnote{Communication between the author and Patrick Clancy, Senior Manager for Strategic Growth, March 8, 2021.}

Since at least early June 2021, Armanino has resumed attestation reports of True USD.  The new owner of is Techteryx, Ltd.  based in Shenzhen.  As for collateral, the audit shows\footnote{\url{https://real-time-attest.trustexplorer.io/truecurrencies}} the balances are held in five banks, including First Digital (61\%) and Signature Bank (17\%).  One hundred million dollars is in unnamed financial institutions.

\subsection{Paxos}

Paxos comes in two flavors, the standard and the gold backed version.  I discuss the standard Pax here.  Paxos Trust Company has engaged Withum,\footnote{\url{https://www.withum.com/our-locations/}} a New Jersey based auditing firm, to provide monthly statements about tokens outstanding and the collateral holdings.  On August 24, 2021, Paxos Standard was rebranded\footnote{\url{https://www.paxos.com/the-digital-dollar-that-always-equals-a-dollar-paxos-standard-pax-is-now-pax-dollar-usdp/}} as Pax Dollar (USDP).

In July 2022, Paxos began to release monthly reports\footnote{\url{https://paxos.com/usdp-transparency/}} with CUSIP level detail on their USDP collateral.  In their report for August 31, 2022, they disclose holdings of six Treasury bills, two repurchase agreements on Treasury bonds, and cash deposits.  They list a set of banks where the cash deposits might be held.

\subsection{Huobi}

The Huobi USD is a stable coin pegged to the U.S. dollar and backed by USD in reserve accounts.\footnote{\url{https://www.huobi.com/en-us/usd-deposit/}}  Attestation reports are provided by Eide Bailly, which reports that the company is based in the British Virgin Islands. As of December 2021, going to the Huobi USD website comes with this warning:  ``Currently, individual / institutional clients from Mainland China, the United States, Iraq, Cuba, Iran, Sudan, Syria, Bangladesh, Ecuador, Tunisia, Libya, Venezuela, etc. are not available to use this service."

\subsection{Terra USD}

Terra USD is a USD pegged stablecoin, collateralized by a digital coin Luna.  It is algorithmically stabilized, requiring destruction of one dollar of Luna for every newly minted Terra USD token.\footnote{Do Kwon, ``Announcing TerraUSD (UST)— the Interchain Stablecoin, \url{https://medium.com/terra-money/announcing-terrausd-ust-the-interchain-stablecoin-53eab0f8f0ac}, September 21, 2020.}

Luna reached its highest value on April 4, 2022, closing at \$116.41. 
During a sharp selloff in digital assets in the Spring of 2022, Luna fell below \$1.00 on May 12.  Terra USD lost its peg to the dollar, falling below \$0.10 by the end of the month.\footnote{\url{https://www.wsj.com/articles/terrausd-crash-led-to-vanished-savings-shattered-dreams-11653649201}} On February 16, 2023, the SEC charged Singapore-based Terraform Labs and founder Do Hyeong Kwon with securities fraud.\footnote{\url{https://www.sec.gov/news/press-release/2023-32}}.

Terra, rebranded as Terra Classic, is trading in mid-February 2023 at \$0.0284.\footnote{\url{https://coinmarketcap.com/currencies/terrausd/}}

\subsection{Gemini Dollar}

Gemini may be best known for its founders, the Winklevoss brothers.  Gemini is on the green list of the New York State Department of Financial Services, and the Gemini Dollar is audited monthly by BPM LLP.\footnote{\url{https://www.gemini.com/dollar}}    Gemini claims that ``Each GUSD corresponds to a U.S. dollar held by Gemini in accounts at U.S. FDIC-insured bank accounts and money market funds holding short-term U.S. treasury bonds and maintained at a custodian.  The pass-through of insurance from the FDIC secured collateral to the token holders is still just a conjecture.\footnote{Shearman \& Sterling LLP, US Stablecoin Regulation: Bringing Stablecoins Into the Regulatory Fold, \url{https://www.jdsupra.com/legalnews/us-stablecoin-regulation-bringing-5449329/}}

Gemini has the most transparent disclosure\footnote{\url{https://assets.ctfassets.net/jg6lo9a2ukvr/VOtyB4tBb0G4FVt6EqFnm/8dae03169f391431130e2e0e73913638/Gemini_Dollar_Examination_Report_03-31-22.pdf}} of where it holds its' assets: ``The Gemini dollar accounts are held and maintained by State Street Bank and Trust Company, Signature Bank, and within a money market fund managed by Goldman Sachs Asset Management, invested only in U.S. Treasury Obligations."

\subsection{Synth sUSD}

sUSD is a token made by the Synthetix team\footnote{\url{https://synthetix.exchange/}} which tracks the price of USD.  Users provide collateral, either Ethereum or Synthetix Network Tokens, when creating sUSD.\footnote{\url{https://docs.synthetix.io/litepaper/}}

Synthetix requires all synthetic tokens, or synths for short, be backed at a collateralization ratio (C-Ratio) of 300\%. Falling below 150\% will trigger a liquidation.\footnote{\url{https://blog.synthetix.io/new-liquidation-mechanism/}}  Synthetix reported an aggregate collateralization ratio of 284\% on May 17, 2022.\footnote{\url{https://stats.synthetix.io/}}

\vspace{0.15in}
Summarizing, USD Coin, Binance USD, Paxos, True USD, and Gemini USD are 100\% backed by cash equivalents. Tether holds 79\%.  Huobi USD claims to be completely backed by cash equivalents, but I could not view the attestation reports from the U.S. There are varying degrees of transparency about where the assets are held.  Gemini, Paxos, Binance USD (all) and True USD (91\%) name the banks where deposits are held.\footnote{Tether does not disclose where assets are domiciled. Paolo Ardoino, the Chief Technology officer of Tether said in a Bloomberg interview ``we are working to expand our banking partners...so we don't want to give away our secret recipe so easily." \url{https://www.bloomberglinea.com/2022/02/17/were-not-revealing-our-secret-recipe-tethers-cto-says/ }} 

Dai and sUSD hold entirely digital assets as collateral.

\subsection{Money Market Mutual Funds}

In this section, I compare the collateral of the stablecoins to money market mutual funds (MMMF) in the United States.  As shown in \autoref{tab:mmmf_assets}, the MMMF have all of their assets in cash equivalents.  To state the obvious, they do not hold any digital assets.
\begin{table}[htbp]
  \centering
  \begin{threeparttable}
  \caption{Asset Holdings of U.S. Money Market Mutual Funds}
    \begin{tabular}{lrr}
    \hline
          \multicolumn{1}{l}{Asset Category} & \multicolumn{1}{l}{Govt.} & \multicolumn{1}{l}{Prime} \\
          \hline \hline
    US Treasury debt & 40.08\% & 0.87\% \\
    US Government agency debt & 8.78\% & 0.07\% \\
    Repurchase agreement & 50.85\% & 28.26\% \\
    Certificates of deposit & 0.00\% & 21.19\% \\
    Non-negotiable time deposit & 0.00\% & 13.11\% \\
    Commercial paper & 0.00\% & 31.61\% \\
    Municipal debt & 0.15\% & 2.46\% \\
    Other & 0.15\% & 2.43\% \\
    \hline
          &       &  \\
    Total Assets (\$bn) & \$4,074.34 & \$423.19 \\
    \hline
    \end{tabular}%
  \label{tab:mmmf_assets}%
\begin{tablenotes}
      \small
      \item Asset holdings as of March 31, 2022.  The data are collected from form SEC N-MFP by the Investment Company Institute.
    \end{tablenotes}
    \end{threeparttable}%
\end{table}%

The prime funds hold commercial paper, time deposits, and certificates of deposit while the U.S. government funds hold primarily U.S. Treasury debt and U.S. Agency and Treasury repo.  Each funds trustee, management, and fee structure are clearly disclosed.  While five stablecoins have a collateral structure similar to U.S. government MMMFs, only Gemini discloses completely its asset structure and location of the collateral.

\subsection{Tests of daily stablecoin volatility}

\citet{LVStable} make the analogy between stablecoin projects and currency pegs.  They argue that even well collateralized stablecoins may face instability.  \citet{GrobysaStability} find that lagged Bitcoin volatility
Granger-causes stablecoin volatility.  \citet{HoangStability} show that at the  5-minute frequency, stablecoin returns are as volatile as Bitcoin.  They also find that stablecoin volume and volatility are highly correlated with BTC, even at the daily frequency.  \citet{Klages} show theoretically that stablecoin price variation, both to the upside and downside, is distinctly greater during deleveraging cycles.

I test formally whether the stablecoins holding cash equivalents are more stable than those holding digital assets.  I compute the daily range for Dai, Terra USD and sUSD for each day between November 25, 2020 and May 16, 2022.  I then take the minima of the daily range of the three currencies.

I then test the relative stability of USD Coin, Tether, Binance USD, True USD, Gemini and Paxos comparing each day's high low to the algorithmic stabilized tokens.  I then use the Wilcoxon test to compare the daily differentials of the trading ranges in \autoref{tab:range_test} over two sub-samples.  

\begin{table}[htbp]
  \centering
  \begin{threeparttable}
  \caption{Range Test for Cash Collateralized Stablecoins}
    \begin{tabular}{lrrrrr}
\hline
          & \multicolumn{1}{l}{2020-11-25 to:} & 2022-03-31 &       & \multicolumn{1}{l}{2022-04-01 to:} & \multicolumn{1}{l}{2022-05-16} \\
    Token & \multicolumn{1}{l}{Range Diff.} & \multicolumn{1}{l}{p-value} &       & \multicolumn{1}{l}{Range Diff.} & \multicolumn{1}{l}{p-value} \\
\hline \hline
    USD Coin & -\$0.0010 & 0.0000 &       & -\$0.0005 & 0.0000 \\
    Tether & -\$0.0013 & 0.0000 &       & -\$0.0015 & 0.0000 \\
    Binance USD & -\$0.0010 & 0.0000 &       & \$0.0020 & 1.0000 \\
    Paxos USD & \$0.0003 & 1.0000 &       & \$0.0053 & 1.0000 \\
    Gemini USD & \$0.0077 & 1.0000 &       & \$0.0066 & 1.0000 \\
    True USD & -\$0.0004 & 0.0000 &       & -\$0.0001 & 0.0227 \\
\hline
    \end{tabular}%
  \label{tab:range_test}%
\begin{tablenotes}
      \small
      \item Each day's range differential is the median difference between the high and low for the stablecoin and the minimum range among the three digital asset stablecoins, Dai, Terra USD and sUSD. A negative value indicates the range is smaller for the cash collateralized token. The test statistic is the one-sided Wilcoxon signed rank test, which is asymptotically chi-squared.
    \end{tablenotes}
    \end{threeparttable}%
\end{table}%

The second sample, April 1 to May 17, 2022 coincides with the extreme pressure on the stablecoin market that caused Terra USD to lose its peg.

The cash collateralized tokens with market capitalizations over one billion USD, USD Coin, Binance USD, Tether and True USD were more stable than the digitally backed tokens prior to this crisis period.  Even before the crisis, Paxos and Gemini had larger intra-daily ranges.

Despite the pressure on the stablecoin market, USD Coin and True USD had median intra-daily ranges that were smaller than in the pre-crisis period.  Aided by an \$8 billion decline in its' outstanding supply, Tether still has a statistically significantly smaller trading range the the digital asset group.  \citet{Jarno}, using non-parametric tests for a sample ending in September 2019, also find that USD Coin is among the most stable ERC-20 tokens.  Extending the sample reverses their negative conclusions about Tether.

These results expand upon \citet{Briola} analysis of Terra Luna's collapse and strengthens their conclusion that algorithmic stablecoins are intrinsically unstable.  I 

\section{Survival Analysis}

The analysis of digital assets has been influenced by the spectacular returns of a handful of successful assets, most notably Bitcoin (BTC).

I construct a complete set of token transactions on the Mainnet.  Between 2016 and the first quarter of 2022, there are more than 392,000 distinct tokens on the network.

\begin{figure}[H]
	\centering
		\caption{Ethereum Active Digital Tokens Outstanding}
		\label{fig:digital_outstanding}
        \begin{minipage}{0.97\linewidth}
        \begin{center}
			\includegraphics[width=0.97\textwidth]{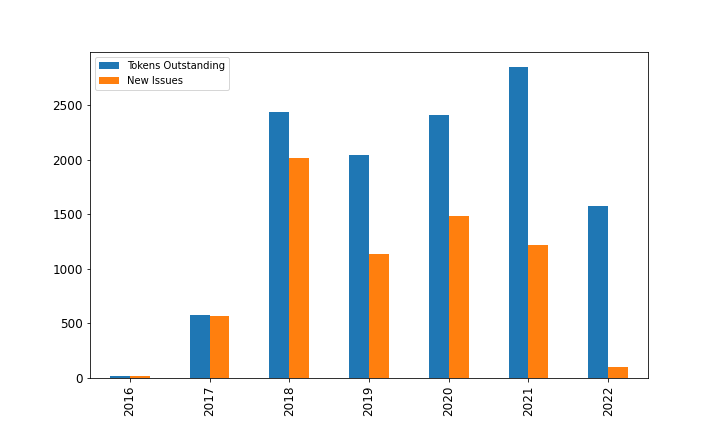} \\
		\end{center}
		\small
		\textit{Note: } Issue dates are based on the token's first transactions on the Mainnet.  I only include tokens with at least 10,000 cumulative transactions.
    \end{minipage}
\end{figure}

To focus the analysis on the active tokens, I only include the 6,558 tokens with at least 10,000 transactions. I graph the growth in \autoref{fig:digital_outstanding}.

\citet{GortonWildcat} compare these developments to the free banking era that began in 1830s.  While Gorton and Zhang claim that private currencies did not contribute to banking panics, they do concede that "varying
discounts made actual transactions (and legal contracting) very difficult...There was constant haggling and arguing over the value of notes in transactions."  The National Bank Act of 1863, they note, effectively ended the era of privately issued bank notes.

\begin{figure}[H]
	\centering
		\caption{Hazard Function for Ethereum Tokens}
		\label{fig:hazard_function}
        \begin{minipage}{0.97\linewidth}
        \begin{center}
			\includegraphics[width=0.97\textwidth]{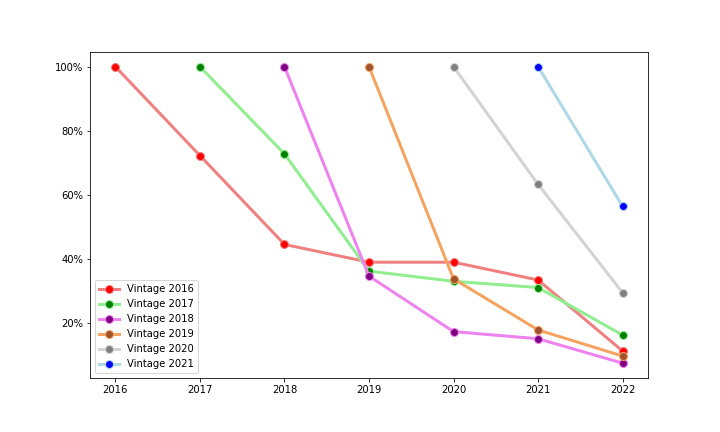} \\
		\end{center}
		\small
		\textit{Note: } Token vintage is determined by the first date of trading on the Ethereum Mainnet. I classify a project as dead if volume falls below 1\% of the peak quarterly volume.
    \end{minipage}
\end{figure}

I also find that many tokens have failed, particularly those started in recent years.  I plot a hazard function in \autoref{fig:hazard_function}. 

There were 1,581 active tokens outstanding at the end of the first quarter of 2022. The peak year for issuance was 2018 when more than 2,000 tokens were issued.


With stable coins, the effects of survivorship are much more concerning because they are not designed to provide any capital gains.

\begin{figure}[H]
	\centering
		\caption{Ethereum Stablecoins Outstanding and New Issues}
		\label{fig:stablecoins_outstanding}
        \begin{minipage}{0.97\linewidth}
        \begin{center}
			\includegraphics[width=0.97\textwidth]{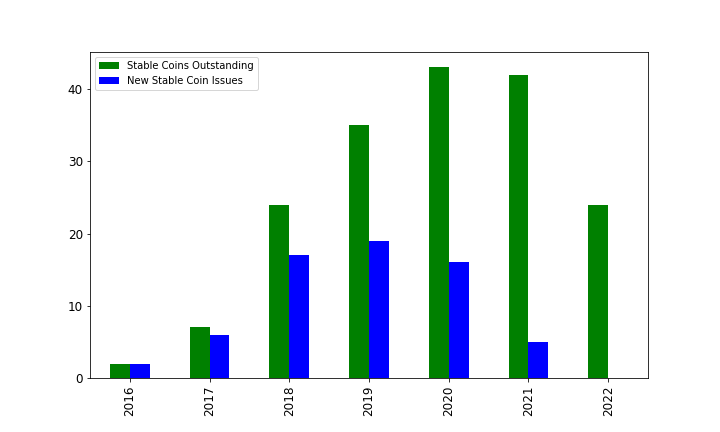} \\
		\end{center}
		\small
		\textit{Note: } The projects are from CoinCodex, CoinMarketCap and Nomics.   The vintage date is from the first transaction on the Mainnet.
    \end{minipage}
\end{figure}

From CoinMarketCap, Nomics, and CoinCodex I identify 91 active and inactive stablecoin projects.  65 of the 91 transact on the Ethereum Mainnet.  These 65 are the most active and include all the stablecoins in the top 5,000 in market capitalization. I graph the growth of Ethereum stablecoins in \autoref{fig:stablecoins_outstanding}.

\begin{figure}[H]
	\centering
		\caption{Stablecoin Hazard Function}
		\label{fig:stablecoins_hazard}
        \begin{minipage}{0.97\linewidth}
        \begin{center}
			\includegraphics[width=0.97\textwidth]{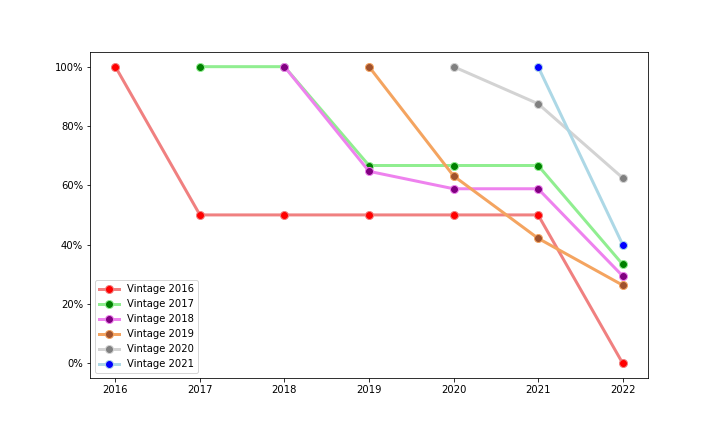} \\
		\end{center}
		\small
		\textit{Note: } The projects are from Blockdata, CoinCodex, CoinMarketCap and Nomics.  The vintage date is from the first transaction on the Mainnet.  I classify a project as dead if volume falls below 1\% of the peak quarterly volume.
    \end{minipage}
\end{figure}

I plot survivor probability in \autoref{fig:stablecoins_hazard} for the 65 ERC-20 stable coins on the Mainnet.  The vintage is the first year that the token transacts on the Mainnet.

DigixDao (DGD) and Xaurum (XAU) are the failures from the 2016 vintage.  Both were backed by gold reserves.  There were six tokens in the 2017 vintage.  Two are still active:  Italian Lira and Tether. There were 17 coins in the 2018 vintage including Paxos, USD Coin, and Gemini from our top ten.  There are now 12 failures, including Pecunio (PCO), and TrueVND (TVND).

2019 was the peak year of new tokens with 19.   Binance USD, Dai, Huobi and True USD, in our top ten, launched that year.  The only other survivor is Ampleforth. Issuance slowed in 2020, with sixteen new stablecoins.  sUSD and  Terra USD are part of that vintage.  The others still trading actively include Carbon and Reserve.  Six tokens have already failed from the 2020 vintage including Stronghold (SHX) and Xank (XANK).  The 2021 vintage has five tokens, only two of which have survived into 2022, Fei USD (Fei) and Liquidity USD (LUSD).

In total, of the 65 major stablecoins that reached the Mainnet between 2016 and 2021, only 24 are still active, a failure rate of 63\%.

\citet{JalanCryptoGold} study the gold based stablecoins.  The authors find that while gold was stable during the COVID crisis, the gold-backed digital assets were as unstable as Bitcoin, despite have no formal or statistical linkage to the leading digital asset.

The hazard functions understate the failure rates of stable coins since many projects never move past the initial offering stage.\footnote{\citet{Phua} found that over 40\% of initial coin offerings were scams, with losses approaching \$12 billion.}  Using data from \citet{Blockreport}, I can identify 103 projects from June 2019.  At the time, 67 projects were active, with more than 30 projects under development.  One of those projects, still not live, is the Libra coin proposed by Facebook.  Libra has been rebranded as Diem and has plans to ``... support single-currency stablecoins (e.g., USD, EUR, and GBP) and a multi-currency coin (XDX)."\footnote{\url{https://www.diem.com/en-us/vision/}}  Facebook has launched a digital wallet called Novi that currently operates with Pax Dollar (USDP).\footnote{\url{https://www.novi.com/}}

I compare statistically the survivor functions for all tokens versus the ERC-20 stablecoins in \autoref{tab:log_rank}.  For the 6,558 active tokens (those which have more than 10,000 cumulative transactions) that enter the Mainnet between 2016 and 2021 and exit prior to March 31, 2022, they survive for an average of 760 days.  For tokens with more than 100,000 transactions, the average survival is 767 days.  The stablecoins survive for an average of 875 days.  A log-rank test fails to reject that the survivor curve of the active tokens is statistically different than the stablecoins.

\begin{table}[htbp]
  \centering
  \begin{threeparttable}
  \caption{Comparison of Survivorship}
    \begin{tabular}{lrrrrr}
\hline
        \multicolumn{1}{l}{Category}  & \multicolumn{1}{l}{\# Trans.Filter} & \multicolumn{1}{l}{\# Tokens} & \multicolumn{1}{l}{Expected Life (Days)} & \multicolumn{1}{l}{Log-Rank Test} & \multicolumn{1}{l}{p-value} \\
\hline \hline
    Tokens & \multicolumn{1}{l}{$>$10,000} & 6,558 & 760   & 12.80 & 0.005 \\
          & \multicolumn{1}{l}{$>$100,000} & 1,096 & 767   & 1.40  & 0.240 \\
          &       &       &       &       &  \\
    Stablecoins &       & 65    & 875   &       &  \\
\hline
    \end{tabular}%
  \label{tab:log_rank}%
\begin{tablenotes}
      \small
      \item I compare the 65 stablecoins to a larger set of Ethereum Mainnet tokens.  The first group has 10,000 or more transactions, and the second 100,000 or more.  The log rank test is asymptotically chi-squared.
    \end{tablenotes}
    \end{threeparttable}%
\end{table}%

In summary, more than half of the stablecoins that reached the Mainnet in the last five years have failed.  This failure rate is not statistically different than other active tokens.

It is important to remember, as I turn to document the rapid growth of the stablecoin sector, that we are studying the survivors.

\section{Transaction Volume}

I report quarterly estimates in \autoref{tab:stable_transactions} of transactions in each of the top ten stable coins for the first quarter of 2022.

\begin{table}[H]
  \centering
  \begin{threeparttable}
  \caption{Transactions Volumes for Top Ten Stablecoins}
\begin{tabular}{lrrr}
\hline
      Token Name &  Transaction Volume (\$bn) &  No. Transactions (mn) \\
\hline \hline
USD Coin & \$704.84 & 4.993 \\
    Tether USD & \$471.47 & 7.308 \\
    Dai Stablecoin & \$154.88 & 0.822 \\
    Binance USD & \$75.19 & 0.094 \\
    Terra USD & \$38.21 & 0.207 \\
    TrueUSD & \$6.81 & 0.034 \\
    Gemini Dollar & \$5.79 & 0.055 \\
    Synth sUSD & \$5.12 & 0.018 \\
    Paxos Standard & \$4.74 & 0.026 \\
    Huobi USD & \$3.29 & 0.004 \\
\hline
\end{tabular}
  \label{tab:stable_transactions}
     \begin{tablenotes}
      \small
      \item The totals are for Ethereum network transactions in 2022Q1.
    \end{tablenotes}
    \end{threeparttable}%
\end{table}%

There were \$1.470 trillion in stablecoin transactions in the first quarter of 2022, up 25.4\% from the same quarter in 2021.  This is almost \$300 billion below the peak transaction volume of \$1.770 trillion in the second quarter of 2021 though.

USD Coin 47.9\% market share, Tether has a 32.1\% market share, , Dai 10.5\%. All of the other coins have less than a 1\% market share.

\begin{figure}[H]
	\centering
		\caption{Stablecoin Quarterly Transaction Volume}
		\label{fig:stablecoin_trans_volume}
        \begin{minipage}{0.97\linewidth}
        \begin{center}
			\includegraphics[width=0.97\textwidth]{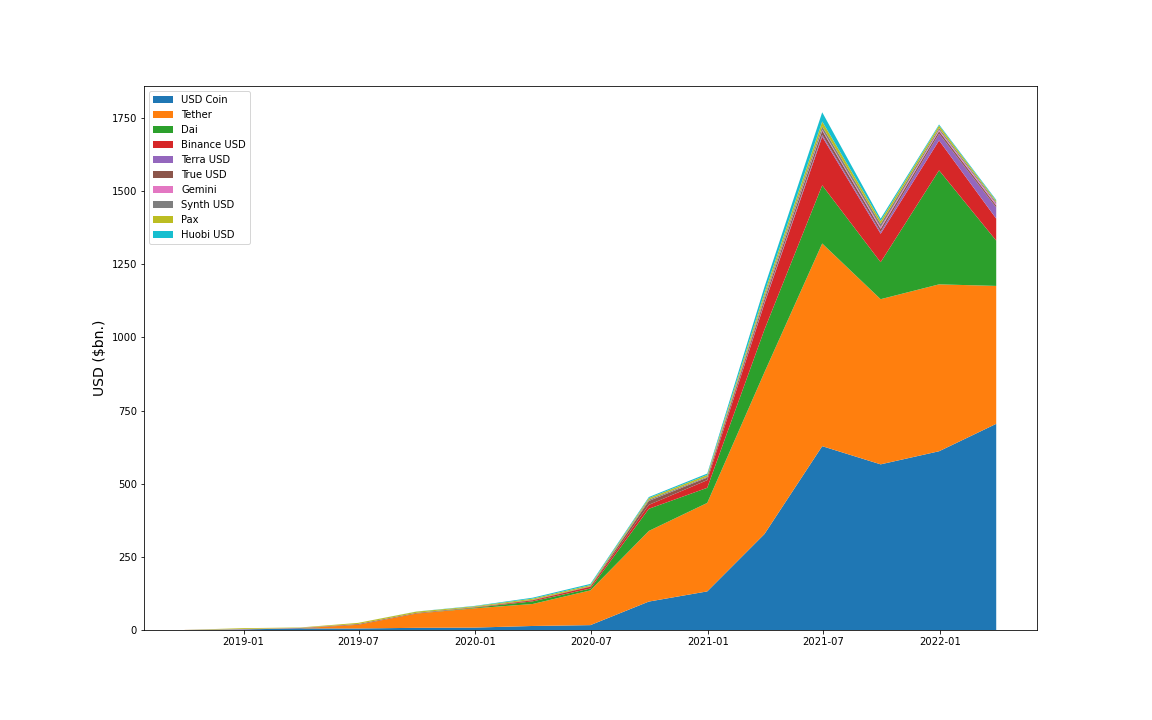} \\
		\end{center}
		\small
		\textit{Note: } The data are from the Ethereum Mainnet The legend is in order of 2022Q1 market share.
    \end{minipage}
\end{figure}

I graph the growth of stablecoin transactions in \autoref{fig:stablecoin_trans_volume}.

The fastest growth rate was for Terra USD, 949\%.  USD Coin grew 114\%, passing Tether in the third quarter of 2021.

\subsection{Velocity}

\begin{figure}[H]
	\centering
		\caption{Stablecoin Velocity}
		\label{fig:stablecoin_velocity}
        \begin{minipage}{0.97\linewidth}
        \begin{center}
			\includegraphics[width=0.97\textwidth]{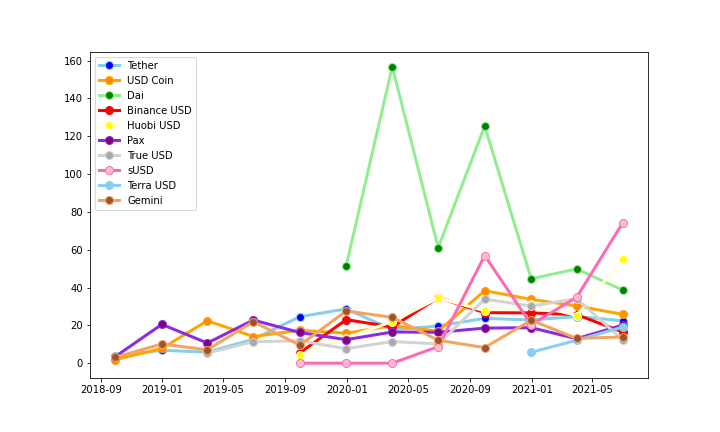} \\
		\end{center}
		\small
		\textit{Note: } The data on supply is just for tokens circulating on the Ethereum Mainnet.
    \end{minipage}
\end{figure}

I take the transaction volume and divide by circulating supply over time.  I have this information from the Mainnet which I compute at the last block for the end of the quarter. I plot velocity in \autoref{fig:stablecoin_velocity}.

Dai has the highest velocity among the three leading stablecoins, 16.8.  USD Coin is second at 15.4, and Tether is third, at 11.8.  The three token (unweighted) average is 14.7.  This average is down 43\% from a year ago.  By comparison, the M1 velocity in 2019Q4 was 5.5, the last quarter not impacted by Covid-19.

\subsection{Trading volume off-chain on centralized exchanges}

The Mainnet had an average transaction time of 13.19 seconds per block\footnote{\url{https://etherscan.io/chart/blocktime}} in 2022Q1 which is far too slow for low latency trading activity.  This largely explains why the vast majority of trading volume in stablecoins occurs off-chain on centralized exchanges.\footnote{Proof-of-stake networks like Solana have achieved latencies under a second, and they are attracting a growing share of the decentralized exchange activity.}   

I study the three most active stablecoins, Tether, USD Coin and Dai.  Each has  with market capitalization over \$9 billion, \$1.5 billion in daily transfer volume on-chain, and more than \$300 million in daily turnover off-chain.\footnote{Binance USD has more than \$8 billion in average daily volume across more than 2,000 pairs, but nearly all of the trading occurs on the Binance exchange.}

I first provide trading volumes across all major currencies and digital asset cross rates in \autoref{tab:volume_all_pairs}.  Tether has more than 24,000 market pairs in which there is a combined average daily volume of nearly \$60 USD billion.  This is comparable to the average daily turnover in New York Stock Exchange listed securities.  USD Coin and Dai trade a combined \$4 billion per day, across many fewer trading pairs.

\begin{table}[htbp]
  \centering
  \begin{threeparttable}
  \caption{Most Active Stablecoins - All Currency Pairs} \label{tab:volume_all_pairs}%
\begin{tabular}{llrr}
\hline
    Name & Symbol & No. Market Pairs &     Avg. Daily Volume (USD bn.)  \\
\hline \hline
    Tether & USDT  & 24,582 & 59.22 \\
    USD Coin & USDC  & 2,656 & 3.94 \\
    Dai   & DAI   & 446   & 0.38 \\
\hline
\end{tabular}
\begin{tablenotes}
\item The table reports the average daily volume
across all currency pairs and centralized exchanges in the stablecoin for the first quarter of 2022.  Source: CoinMarketCap.  The number of trading pairs is from nomics.com
  \end{tablenotes}
  \end{threeparttable}
\end{table}%

I just want to re-iterate that these transactions are \textit{not} part of the transfer volume studied earlier.  From the perspective of the blockchain, these trades are internal to a single wallet and not reported on the Mainnet.

I study the five most active centralized digital asset exchanges. These exchanges have more than 50\% of the market share for the USD stable currencies I study.  I report market shares in \autoref{tab:exchange_volume_market_shares}.

\begin{table}[htbp]
  \centering
  \begin{threeparttable}
  \caption{USD Stablecoin Exchange Market Shares}  \label{tab:exchange_volume_market_shares}%
    \begin{tabular}{lrrlll}
\hline
          & \multicolumn{1}{l}{Binance} & \multicolumn{1}{l}{Bitfinex} & Coinbase & Huobi & Kraken \\
\hline \hline
    Tether & 84\%  & 1.4\% & 1.4\%     & 12\% & 1.7\%  \\
    USD Coin & 72\%  & {$<1\%$} & 13\% & 1.5\%  & 13\% \\
    Dai   & 67\% & {$<1\%$} & 18\% & 1.6\%  & 13\%  \\
\hline
    \end{tabular}%
\begin{tablenotes}
\item The table reports average daily market shares
across all currency pairs in the stablecoin on the exchange for the first quarter of 2022.  Source: CryptoCompare API.
  \end{tablenotes}
  \end{threeparttable}
\end{table}%
\pagebreak

\begin{figure*}
\caption{Trading Volume by Exchange of Major Stablecoins}
\label{fig:stablecoin_exchange_volume}
\begin{multicols}{2}
\center (a) Tether
\includegraphics[width=\linewidth]{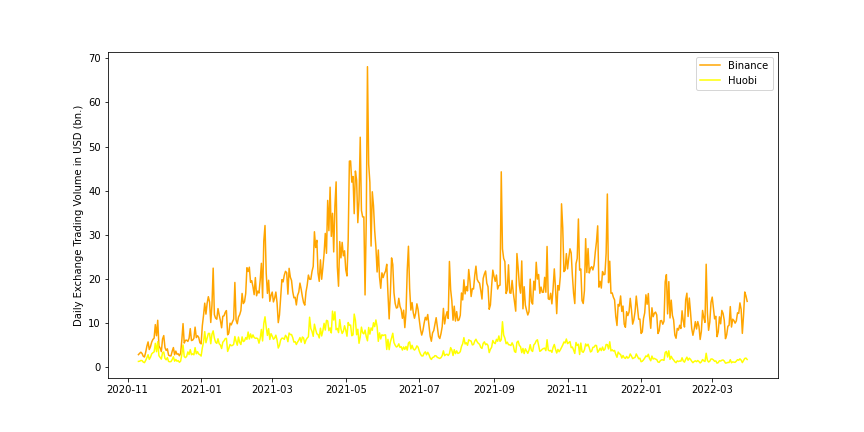}
\center (b) USD Coin
\includegraphics[width=\linewidth]{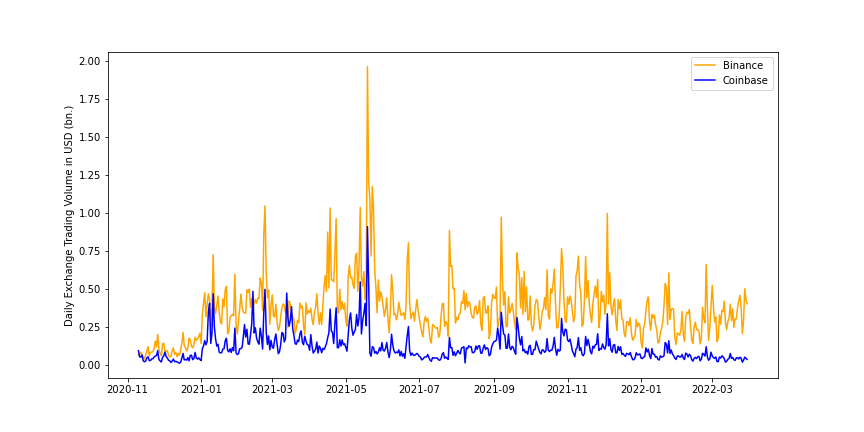}\par
\end{multicols}
\center (c) Dai \\
\includegraphics[width=0.5\linewidth]{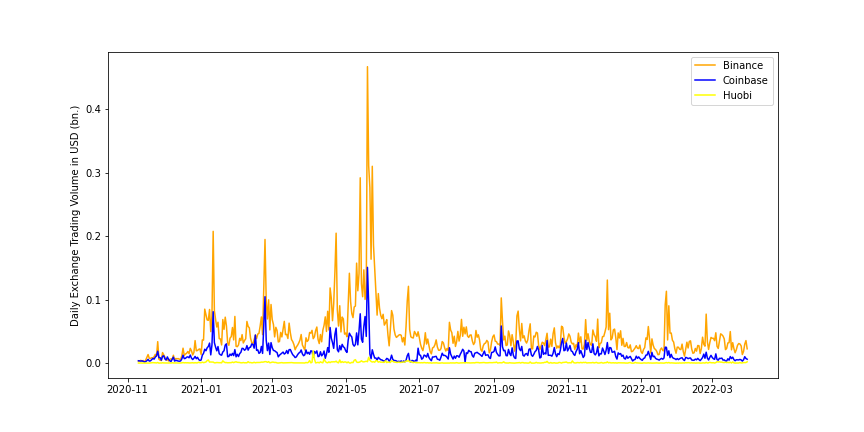}\par
\justify
\textit{Note:} Volumes (in billions of USD) are from the CryptoCompare API for all currency pairs in the stablecoin. 
\end{figure*}

I report the daily trading volume for the three major currencies that trade across multiple exchanges in  \autoref{fig:stablecoin_exchange_volume}.

Binance is the most active exchange in Tether, with a daily average volume of \$11.5 billion in the first quarter of 2022.  The only other exchange-currency pair with over one billion in daily volume is Tether on Huobi (\$1.6 billion daily average). USD Coin on Binance at \$303 million has the third largest average daily volume.

The exchanges do occasionally handle much larger volumes.  Binance trades more than \$68 billion in Tether on May 19, 2021.  All of the other stablecoins have major spikes on that date across all exchanges.

\section{Network Structure}

The Biden Administration has recommended that stablecoin issuers become insured depository institutions.  In this section, I compare the asset concentration of leading stablecoins to comparable banking intermediaries.

\begin{figure}[H]
	\centering
		\caption{Stablecoin Network Size}
		\label{fig:stablecoin_addresses}
        \begin{minipage}{0.97\linewidth}
        \begin{center}
			\includegraphics[width=0.97\textwidth]{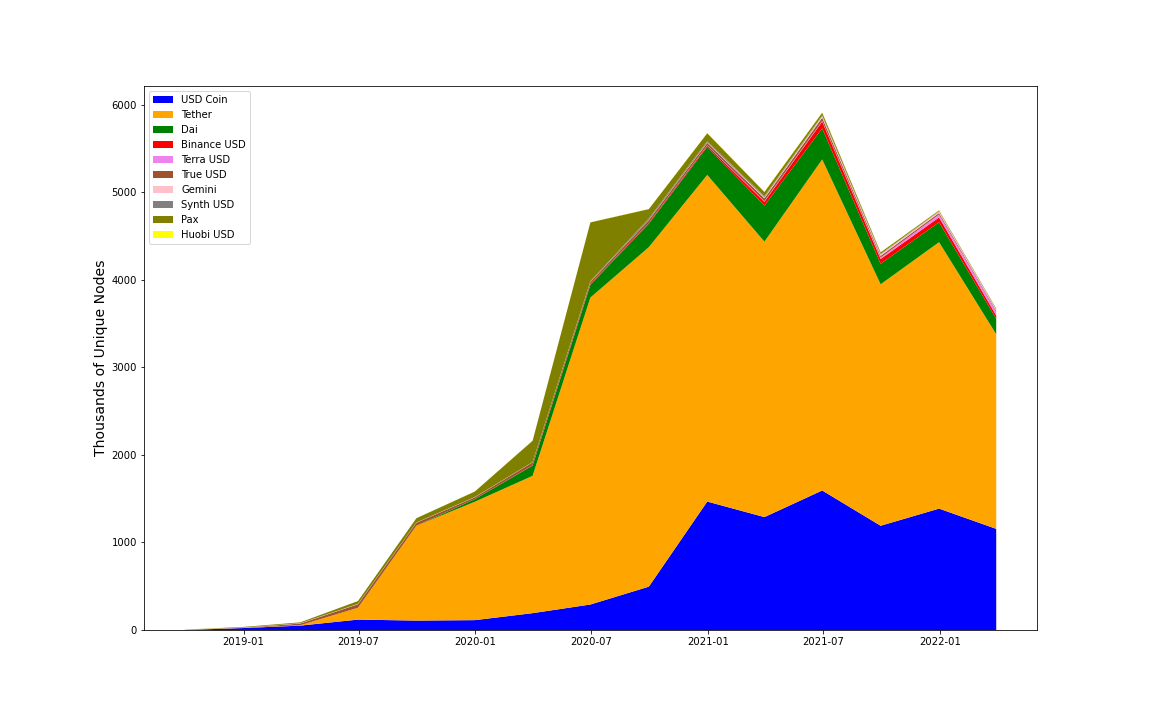} \\
		\end{center}
		\small
		\textit{Note: } The data are from the Ethereum Mainnet and include wallets, smart contracts, and internal transfers.
    \end{minipage}
\end{figure}

The Mainnet provides counterparty information from the transaction initiator and recipients.  I define a counterparty as a wallet or smart contract and include internal transfers in the totals.  I calculate the number of unique counterparties each quarter for each of the tokens.  The time series is plotted in \autoref{fig:stablecoin_addresses}.

In the first quarter of 2022, Tether transactions involve 2.2 million unique counterparties.  USD Coin has 1.16 million.  Dai has 177,000.  The remaining tokens have fewer than 50,000 counterparties.

The Federal Deposit Insurance Corporation estimates that 124.2 million households had access to a bank account.\footnote{Federal Deposit Insurance Corporation, ``How America Banks:
Household Use of Banking and Financial Services," 2020. \url{https://www.fdic.gov/analysis/household-survey/}} Stablecoins are clearly still a niche.  I next analyze the most frequent participants in the Ethereum stablecoin network.

\subsection{Most active addresses}

I provide more detail on the most active network addresses.  The vast majority of transfers occur into or out of exchanges.  I show the top ten addresses for Tether, USD Coin and Dai in \autoref{tab:transfer_addresses}.

\begin{table}[htbp]
  \centering
  \begin{threeparttable}
  \caption{Most Active Transfer Addresses}  \label{tab:transfer_addresses}
    \begin{tabular}{rrrrrrlr}
    \hline
    USDT  &       &       & USDC  &       &       & \multicolumn{1}{r}{DAI} &  \\
    \multicolumn{1}{l}{Address} & \multicolumn{1}{l}{Share} &       & \multicolumn{1}{l}{Address} & \multicolumn{1}{l}{Share} &       & Address & \multicolumn{1}{l}{Share} \\
    \hline \hline
    \multicolumn{1}{l}{Coinbase*        } & 18.86\% &       & \multicolumn{1}{l}{Coinbase*} & 12.24\% &       & Uniswap V3* & 1.27\% \\
    \multicolumn{1}{l}{Binance*         } & 8.65\% &       & \multicolumn{1}{l}{Uniswap V3        } & 6.66\% &       & Coinbase           & 0.96\% \\
    \multicolumn{1}{l}{FTX                } & 5.41\% &       & \multicolumn{1}{l}{FTX               } & 4.33\% &       & Binance            & 0.76\% \\
    \multicolumn{1}{l}{Crypto.com         } & 4.76\% &       & \multicolumn{1}{l}{Uniswap V2*} & 4.29\% &       & Contract 0x74de5…  & 0.58\% \\
    \multicolumn{1}{l}{Address 0xec30d…   } & 4.56\% &       & \multicolumn{1}{l}{Contract 0x74de5… } & 3.75\% &       & Uniswap V2         & 0.47\% \\
    \multicolumn{1}{l}{Uniswap V3         } & 2.96\% &       & \multicolumn{1}{l}{Binance           } & 2.95\% &       & Sushi Swap         & 0.38\% \\
    \multicolumn{1}{l}{OKEx               } & 2.85\% &       & \multicolumn{1}{l}{Circle            } & 2.71\% &       & Contract 0x220bd… & 0.35\% \\
    \multicolumn{1}{l}{Contract 0x74de5d… } & 2.80\% &       & \multicolumn{1}{l}{Sushi Swap        } & 1.59\% &       & Address 0x00000…   & 0.31\% \\
          &       &       &       &       &       & Balancer           & 0.30\% \\
    \hline
    \end{tabular}%
\begin{tablenotes}
\item The table reports market shares of transfer volume by the top ten addresses in the
the first quarter of 2022. * indicates that more than one address is grouped under the name.
  \end{tablenotes}
  \end{threeparttable} 
\end{table}%

There are five centralized exchanges in \autoref{tab:transfer_addresses}: Binance, Coinbase; Crypto.com; FTX; and OKEx; There are two decentralized exchanges, Uniswap\footnote{\url{https://uniswap.org/faq/}: ``Uniswap is a protocol for creating liquidity and trading ERC-20 tokens on Ethereum. It eliminates trusted intermediaries and unnecessary forms of rent extraction, allowing for fast, efficient trading. Where it makes tradeoffs decentralization, censorship resistance, and security are prioritized. Uniswap is open-source software licensed under GPL."}  and Sushi Swap.\footnote{\url{https://docs.sushi.com/}: ``SushiSwap is an automated market-making (AMM) decentralized exchange (DEX) currently on the Ethereum blockchain."} 

The largest decentralized exchange, Uniswap, is the first or second most active set of transfer addresses for USDC and Dai.  Coinbase is the most active in Tether and USD Coin.\footnote{These addresses are the fiat gateways to the exchanges.  Once the assets are on the exchanges, they will no longer report to the Mainnet.}  Binance is the second most active in Tether.  Not all addresses are identified. The fifth largest transfer destination for Tether is an anonymous wallet.

I now turn to see who are the largest holders of the coins.

\subsection{Holders}

The Mainnet allows us to compute the holdings of any ERC-20 token by wallet address. To provide comparison across the top stablecoins, I report the largest holder and Herfindahl (HHI) index for the ten largest stablecoins in \autoref{tab:holding_table}.

\begin{table}[htbp]
  \centering
  \begin{threeparttable}
  \caption{Concentration of Stablecoin Holdings}
    \begin{tabular}{lrrl}
\hline
    Stablecoin & \multicolumn{1}{l}{HHI} & \multicolumn{1}{l}{Largest Share (\%)} & Largest Holder \\
\hline \hline
    BUSD  & 0.408415 & 57.06\% & Binance 8 \\
    GUSD  & 0.295182 & 52.09\% & Gemini 4 \\
    HUSD  & 0.269697 & 48.99\% & Huobi 10 \\
    PAX   & 0.248898 & 49.12\% & 0x7bbd8... \\
    TUSD  & 0.124565 & 28.00\% & Binance 8 \\
    SUSD  & 0.068103 & 17.87\% & Curve.fi: sUSD v2 Swap \\
    UST   & 0.064534 & 19.57\% & Binance 8 \\
    DAI   & 0.041811 & 15.95\% & Curve.fi: DAI/USDC/USDT Pool \\
    USDC  & 0.018301 & 9.68\% & Maker: PSM-USDC-A \\
    USDT  & 0.008754 & 4.51\% & Tether: Treasury \\
\hline
    \end{tabular}%
  \label{tab:holding_table}%
    \begin{tablenotes}
      \small
      \item I retrieve from Etherscan the top 50 holders on April 27, 2022.
    \end{tablenotes}
    \end{threeparttable}
\end{table}%

Binance USD, Gemini, Huobi USD, and Pax USD have the most concentrated holdings.  For Binance USD and Gemini, the largest holder has more than 50\% of the issue.  Both are the exchanges linked to the token, Binance 8 for BUSD and Gemini 4 for GUSD.

I compare my stablecoin HHI estimates to those in the banking industry for context. At the county level, \citet{Meyer} found an average HHI of 0.3468. Binance USD, Gemini, Houbi and Paxos are above that level.  Based on Department of Justice (DOJ) merger guidelines, anything above 0.25 is ``highly concentrated".  Mergers which raise the HHI by 0.02 are also considered worth reviewing.  By DOJ standards, seven of the top ten tokens are ``not concentrated." 

\section{On-Chain Fees}

Transferring ownership of a stablecoin requires intermediaries called \textit{miners} to update the blockchain network.  The fees they earn are transparent to all network participants.

\subsection{Aggregate transfer fees}

A transfer begins with two counterparties, identified by addresses, on the blockchain, with the owner requesting to process a transfer. I will use a real example to help understand the nomenclature:

\begin{table}[htbp]
  \centering
  \caption{Tether Transaction Example}
    \begin{tabular}{llr}
    \hline
    Field & Entry \\
    \hline
    \hline
    Timestamp &  Mar-21-2021 12:00:40 AM +UTC &  \\
    From  &  0xb3eb794a375d802876f67f59d5494b2078f0bdd8  \\
    To    &  0x32034114ac386374d2f3e3057d61fdc3222c49ee  \\
    Contract &  0xdac17f958d2ee523a2206206994597c13d831ec7 (Tether)\\
    Tokens Transferred & 99.742115 &  \\
    Transaction Hash & 0xd79cf3fea1ca4ddaac0f42f98c496159dbe8be1582c17ffbf146096573a373ef   \\
    Block & 12078834   \\
    Gas Price (Ether Gwei) & 140.4887   \\
    Miner & 0xd224ca0c819e8e97ba0136b3b95ceff503b79f53 (UUPool) \\
    Transaction Fee (Ether) & 0.0078967293383   \\
    USD/ETH price & \$1,783.94  (at timestamp)   \\
    Fee in USD & \$14.08729134  \\
    \hline
    \end{tabular}%
  \label{tab:tether_trans_example}%
\end{table}%

In the example in \autoref{tab:tether_trans_example}, wallet 0xb3eb794a375d802876f67f59d5494b2078f0bdd8 initiates a transfer of 99.742115 Tether (USDT) to wallet 0x32034114ac386374d2f3e3057d61fdc3222c49ee.  Once the transaction begins, a hash number is assigned to the transfer.  All of the information in the example is public to the network, and I have retrieved it from the Mainnet.

The transfer request is publicized on the network and miners must make the effort to insert the transaction into the blockchain.  The miners compete and pricing varies considerably with network congestion.

The transfer was included in Block 12078834.  The miner UUPool completed the block in forty seconds.  It charged a fee (priced in ether) using what is called the gas price.  This name reflects the fact that the miner must use computer time to incorporate the transfer into the blockchain.  

Gas prices are typically quoted in Gwei, billions of wei, the smallest divisible unit of ether.  The gas price for this transaction was 140.4887 Gwei, or 140.4887 $\times10^{-9}$ in ether terms.  The transaction used 56,209 units of gas, making the ether cost of the transaction 7.90$\times10^{-3}$ ether.

\begin{figure}[H]
	\centering
		\caption{Average Ether Gas Prices Jan. 1, 2018-March 31, 2022}
		\label{fig:average_gas}
        \begin{minipage}{0.97\linewidth}
        \begin{center}
			\includegraphics[width=0.97\textwidth]{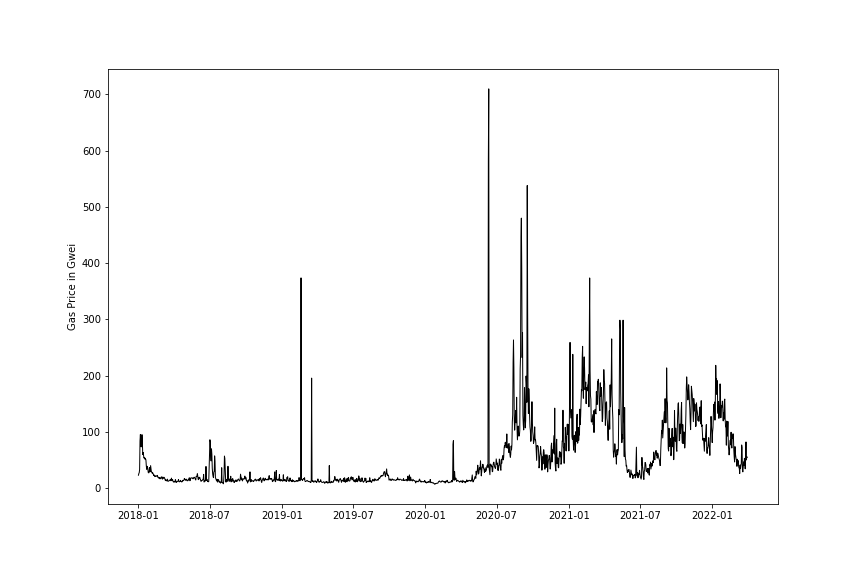} \\
		\end{center}
		\small
		\textit{Note: } The data are from Etherscan, \url{https://etherscan.io/chart/gasprice}.  
    \end{minipage}
\end{figure}

I plot historical daily average ether gas prices in \autoref{fig:average_gas}.

Average prices have ranged from 7.32 to 709.71 Gwei between January 2018 and March 2022.  There is also substantial intra-daily fluctuation not captured in this figure.  On March 14, 2021, for example, gas prices ranged from 70 to 600 Gwei.

The USD price of ether at the timestamp of the transaction completion in \autoref{tab:tether_trans_example} was \$1,783.94, so the dollar cost of the transaction was \$14.09.

\subsubsection{Internal transactions}

Internal transactions are transactions between smart contracts.  ERC-20 tokens are themselves smart contracts, but a chain of token transfers may include numerous internal transfers with additional intermediaries.

I discuss an example in \autoref{tab:internal_trans_table}.  The transaction is two swaps, on the largest decentralized exchange, Uniswap, that result in the exchange of Gemini Dollar for USD Coin.

\begin{table}[htbp]
  \centering
  \caption{Internal Transactions}
    \begin{tabular}{llr}
\hline
    Field & Entry \\
\hline \hline
    Transaction Hash & 0x00085bce0480ffae3a717aa1af9f72cecb7566b2ed8ce4cb50b4e8cc225eebe2 &  \\
    Date  & Jan-19-2021 07:54:01 AM &  \\
    Transaction Action: & & \\
    Swap & 10,623.08 GUSD for 7.87795 Wrapped Ether On Uniswap V2 \\
    Swap & 7.87795 Ether For 10,779.054879 USDC On Uniswap V2 \\
    Gas Price (Ether Gwei): & 132.28    &  \\
    Transaction Fee (Ether): & 0.06074 &  \\
    USD/ETH price: & \$1,367.65 (at timestamp) &   \\
    Total Fee in USD: & \$83.07 & \\
\hline
    \end{tabular}%
  \label{tab:internal_trans_table}%
\end{table}%

In calculating the network size, I include the wallet that supplies Gemini USD to the swap pools, the automated market maker smart contracts, and the wallet receiving the USD Coin.  I will average the gas fee across all of the steps of the transaction.

These assumptions have little impact on the fees for Tether and USDC.  Only 0.2\% of Tether and 1.4\% of USD Coin transactions involve multiple steps like in \autoref{tab:internal_trans_table}.  For Dai, while the number is higher, 10.8\%, it has little effect on the distribution of transaction fees.

\begin{figure}[H]
	\centering
		\caption{Quarterly Transaction Fees for Top Stablecoins}
		\label{fig:quarterly_fees}
        \begin{minipage}{0.97\linewidth}
        \begin{center}
			\includegraphics[width=0.97\textwidth]{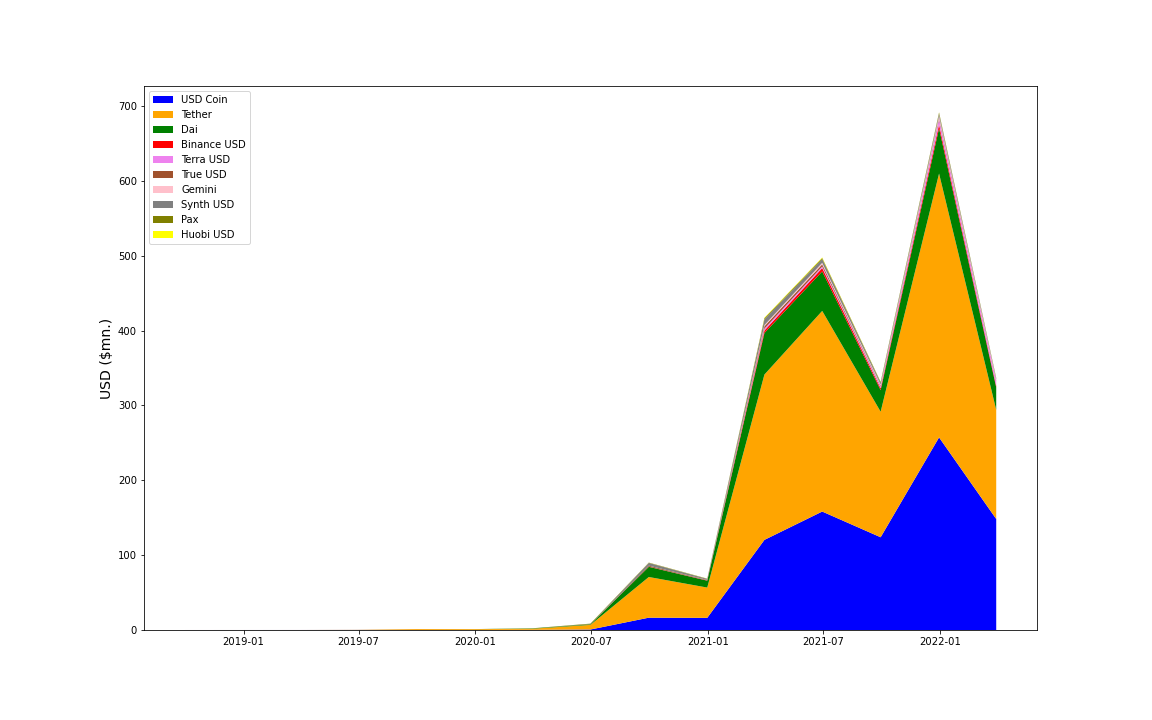} \\
		\end{center}
		\small
		\textit{Note: } I compute fees in Ether from the Mainnet and convert them to USD using daily 12:00 GMT Ethereum prices.
    \end{minipage}
\end{figure}

I aggregate fees over the last three years for all the top ten stablecoins and plot them in \autoref{fig:quarterly_fees} 

The total fees from 2022Q1 are \$336.80 million and are dominated by three coins.  USDC generates \$148 million in fees (44.0\%), USDT \$146 million (43.9\%), and Dai, \$30 million (8.9\%).  The other seven coins only generate \$12.7 million in fees, led by Terra USD  at \$6.1 million.  Fees are down 51.3\% from the peak of \$691.7 million in the fourth quarter of 2021.
 
\subsection{Cross-section of fees}

My examples show that transaction costs can vary substantially. Since block insertion time is influenced by the transfer fee, both costs and time to completion can be hard to predict.  The cost of transfers has also been rising over time because of network congestion and the upward trend in Ether prices.

\subsubsection{Tether fees}

I begin the discussion on fees with Tether because it has the the number of transactions, 7.31 million in the first quarter of 2022.  Summary statistics are in \autoref{tab:usdt_trans_stats}.

\begin{figure}[H]
	\centering
		\caption{USDT Transaction Fee Histogram - USD}
		\label{fig:usdt_fee_hist}
        \begin{minipage}{0.97\linewidth}
        \begin{center}
			\includegraphics[width=0.97\textwidth]{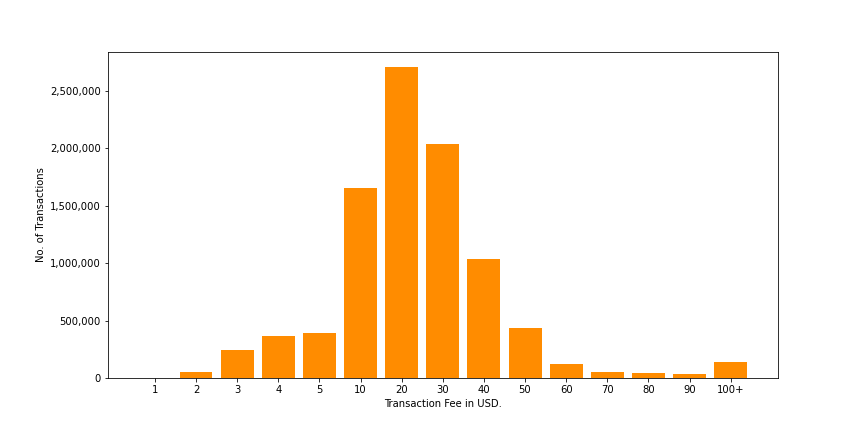} \\
		\end{center}
		\small
		\textit{Note: } The histogram describes transaction fees in USD for all Mainnet transactions in the first quarter of 2022.
    \end{minipage}
\end{figure}

\begin{table}[htbp]
  \centering
  \caption{Tether (USDT) Transactions for 2022Q1}
    \begin{tabular}{lrrrrrrr}
\hline
          & \multicolumn{1}{l}{Mean} & \multicolumn{1}{l}{STD.} & 0.01  & 0.25  & 0.5   & 0.75  & 0.99 \\
\hline \hline
    Token Value & 68,368 & 1,658,282 & 10    & 308   & 1,280 & 7,809 & 999,900 \\
    Gas Price (Gwei) & 92.88 & 127.37 & 16.39 & 42.00 & 75.49 & 122.16 & 328.81 \\
    Gas Used & 76,185 & 91,146 & 41,297 & 46,085 & 58,397 & 63,209 & 404,923 \\
    Fee USD & \$21.14 & \$55.86 & \$2.13  & \$6.77  & \$13.03 & \$23.23 & \$143.28 \\
    Fee Pct. & 7.44\% & 26.24\% & 0.00\% & 0.17\% & 0.93\% & 4.20\% & 179.35\% \\
\hline
    \end{tabular}%
  \label{tab:usdt_trans_stats}%
\end{table}%
The median fee is \$13.03 with a median percentile fee of 0.93\%.  I plot a histogram of fees in USD in \autoref{fig:usdt_fee_hist}. 

Only 4.2\% of fees are under the average \$3.08 fee for out-of-network ATM transactions.\footnote{\url{https://www.bankrate.com/banking/checking/checking-account-survey/}}  There is a long tail as well:  22\% of fees are over \$25 and 2.1\% have fees of over \$100. 

\begin{figure}[H]
	\centering
		\caption{USDT Transaction Fee Histogram - Pct.}
		\label{fig:usdt_fee_hist_pct}
        \begin{minipage}{0.97\linewidth}
        \begin{center}
			\includegraphics[width=0.97\textwidth]{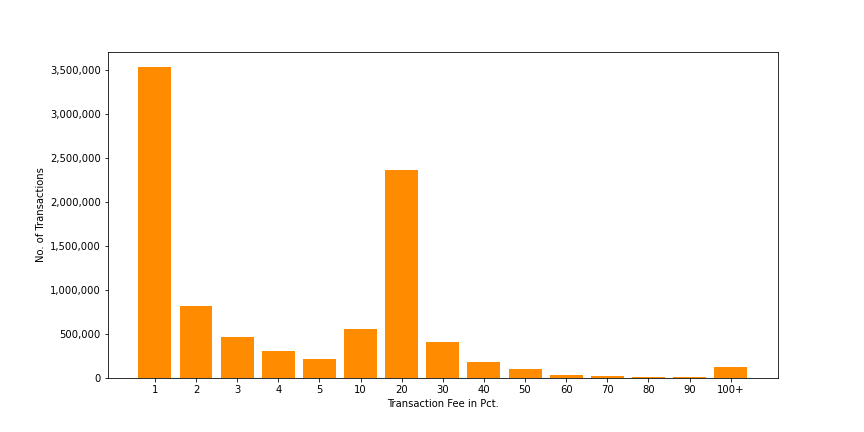} \\
		\end{center}
		\small
		\textit{Note: } The histogram describes fees as a percent of the transferred amount for all Mainnet transactions in the first quarter of 2022.
    \end{minipage}
\end{figure}

The fees in percentage terms are in \autoref{fig:usdt_fee_hist_pct}

While the median fee is under 1\%, there are bad surprises though.  123,065 transactions (1.78\%) have fees which exceed the transferred value.

There are economies of scale.  Transactions over \$10,000 have a median fee of 0.032\%.  Transfers under that amount have a median fee of 1.74\%.

\subsubsection{USD Coin fees}

I analyze all of the 4.99 million USD Coin transactions in the first quarter of 2022.  Summary statistics are in \autoref{tab:usdc_trans_stats}.
Descriptive statistics are in \autoref{tab:usdc_trans_stats}

\begin{table}[htbp]
  \centering
  \caption{USD Coin (USDC) Transactions for 2022Q1}
    \begin{tabular}{lrrrrrrr}
\hline
          & \multicolumn{1}{l}{Mean} & \multicolumn{1}{l}{Std. Dev.} & 0.01  & 0.25  & 0.5   & 0.75  & 0.99 \\
\hline \hline
    Token Value & 166,295 & 2,545,721 & 5     & 540   & 2,963 & 19,950 & 2,416,698 \\
    Gas Price Gwei & 97.06 & 174.98 & 16.54 & 42.94 & 76.53 & 123.91 & 367.00 \\
    Gas Used & 122,800 & 139,582 & 43,713 & 48,537 & 65,625 & 162,302 & 628,296 \\
    Fee USD & \$35.00 & \$98.81 & \$2.59  & \$9.69  & \$18.84 & \$36.51 & \$246.98 \\
    Fee Pct. & 7.62\% & 30.63\% & 0.00\% & 0.11\% & 0.66\% & 3.25\% & 315.66\% \\
\hline 
    \end{tabular}%
  \label{tab:usdc_trans_stats}%
\end{table}%

Transaction costs in USD are higher for USD Coin than Tether, \$18.84 versus \$13.03 for Tether.  The median transaction size for USD Coin is more than twice as large as Tether, so this helps bring USD Coin's median percentile fee of 0.66\% below Tether's 0.93\%.

\begin{figure}[H]
	\centering
		\caption{USDC Transaction Fee Histogram - USD}
		\label{fig:usdc_fee_hist}
        \begin{minipage}{0.97\linewidth}
        \begin{center}
			\includegraphics[width=0.97\textwidth]{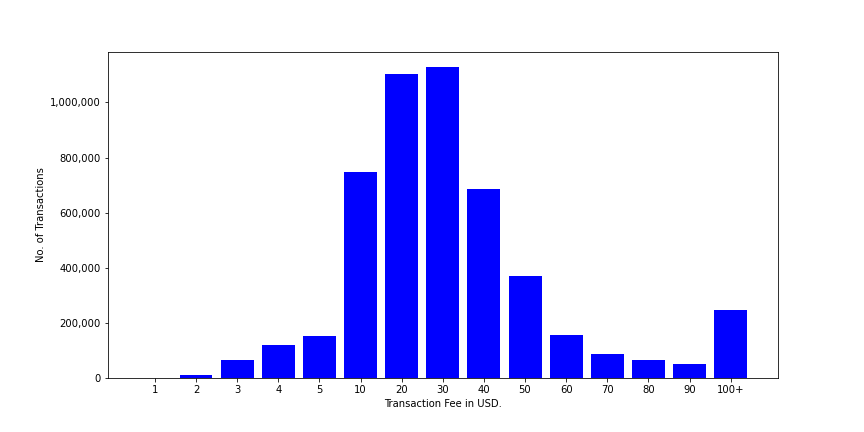} \\
		\end{center}
		\small
		\textit{Note: } The histogram describes transaction fees in USD for all Mainnet transactions in the first quarter of 2022.
    \end{minipage}
\end{figure}

I plot a histogram of fees in USD in \autoref{fig:usdc_fee_hist}.

Only 1.9\% of fees are below the ATM threshold of \$3.00.  More than a third of fees are above \$25.00, and 5.8\% are above \$100.

\begin{figure}[H]
	\centering
		\caption{USDC Transaction Fee Histogram - Pct.}
		\label{fig:usdc_fee_hist_pct}
        \begin{minipage}{0.97\linewidth}
        \begin{center}
			\includegraphics[width=0.97\textwidth]{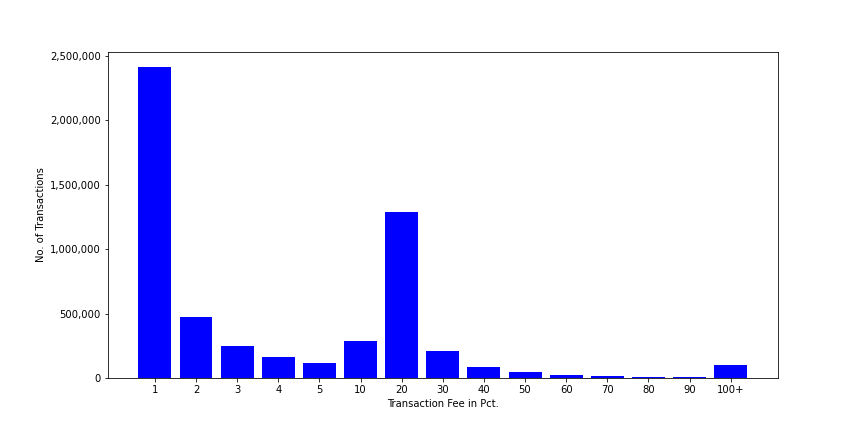} \\
		\end{center}
		\small
		\textit{Note: } The histogram describes fees as a percent of the transferred amount for all Mainnet transactions in the first quarter of 2022.
    \end{minipage}
\end{figure}

The fees in percentage terms are in \autoref{fig:usdc_fee_hist_pct}

100,725 transactions (2.4\%) of fees are larger than the transferred amount.  Bigger trades ($>$\$10,000) pay a median fee of 0.045\%, but smaller trades have a median of 1.74\%.

\subsubsection{Dai Stablecoin}

I analyze all of the 822,000 Dai transactions in the first quarter of 2022.  Descriptive statistics are in \autoref{tab:dai_trans_stats}.

\begin{table}[htbp]
  \centering
  \caption{Dai Stablecoin Transactions for 2022Q1}

    \begin{tabular}{lrrrrrrr}
    \hline
          & \multicolumn{1}{l}{Mean} & \multicolumn{1}{l}{STD.} & 0.01  & 0.25  & 0.5   & 0.75  & 0.99 \\
        \hline \hline
    Token Value & 249,040 & 8,874,678 & 3     & 578   & 4,327 & 30,627 & 2,509,596 \\
    Gas Price Gwei & 96.50 & 346.32 & 15.96 & 40.91 & 73.38 & 120.34 & 374.40 \\
    Gas Used & 167,594 & 238,717 & 29,918 & 50,525 & 108,669 & 195,151 & 1,145,323 \\
    Fee USD & \$48.20 & \$160.23 & \$1.78  & \$9.03  & \$19.80 & \$47.21 & \$438.01 \\
    Fee Pct. & 7.91\% & 32.39\% & 0.00\% & 0.08\% & 0.47\% & 2.99\% & 650.16\% \\
\hline
    \end{tabular}%
  \label{tab:dai_trans_stats}%
\end{table}%

Dai has the largest median transaction size, 4,327 tokens,  of the three major stablecoins.  The median fee in USD is \$19.80, a dollar higher than USD Coin, and six dollars higher than Tether.

\begin{figure}[H]
	\centering
		\caption{USDC Transaction Fee Histogram - USD}
		\label{fig:dai_fee_hist}
        \begin{minipage}{0.97\linewidth}
        \begin{center}
			\includegraphics[width=0.97\textwidth]{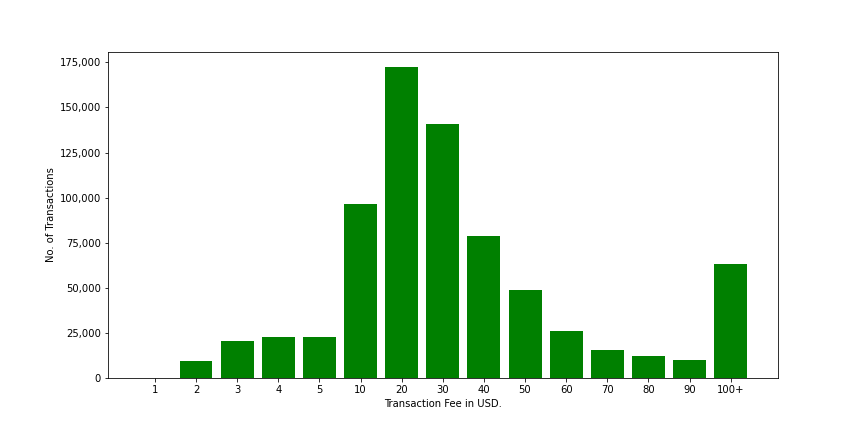} \\
		\end{center}
		\small
		\textit{Note: } The histogram describes transaction fees in USD for all Mainnet transactions in the first quarter of 2022.
    \end{minipage}
\end{figure}

A histogram of fees in USD is in \autoref{fig:dai_fee_hist}.

4.8\% of fees are below the \$3.00 ATM threshold.  More than 42\% are above \$25 and over 10\% are above \$100.

\begin{figure}[H]
	\centering
		\caption{Dai Transaction Fee Histogram - Pct.}
		\label{fig:dai_fee_hist_pct}
        \begin{minipage}{0.97\linewidth}
        \begin{center}
			\includegraphics[width=0.97\textwidth]{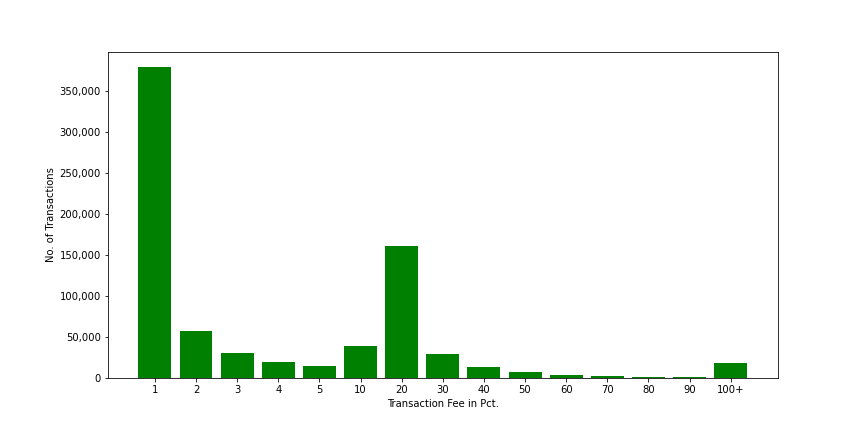} \\
		\end{center}
		\small
		\textit{Note: } The histogram describes fees as a percent of the transferred amount for all Mainnet transactions in the first quarter of 2022.
    \end{minipage}
\end{figure}

The fees in percentage terms are in \autoref{fig:dai_fee_hist_pct}

The median fee in percentage terms is 0.47\% which is lower than Tether and USD Coin, driven by the larger transaction sizes. Bigger transfers ($>$\$10,000) pay lower fees, 0.062\%, but these are still higher than the fees for institutional size trades for the USDT and USDC.  There are 18,024 transfers in which fees exceed the transferred amount.

\subsection{Trends over time}

Ethereum prices rose dramatically in 2021.  They averaged \$1,534.91 in 2021Q1 and \$4,095.57 in 2021Q4, an increase of 267\%.

\begin{figure}[H]
	\centering
		\caption{Transaction Cost Trends in USD}
		\label{fig:trans_fee_USD_trends}
        \begin{minipage}{0.97\linewidth}
        \begin{center}
			\includegraphics[width=0.97\textwidth]{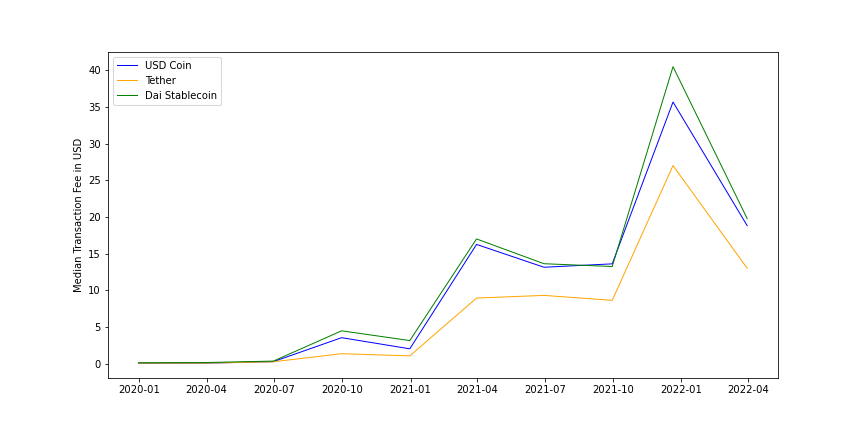} \\
		\end{center}
		\small
		\textit{Note: } I use the entire set of transactions in each quarter for each of the three major tokens and show the rise in median fees, measured in USD per transfer, over time.
    \end{minipage}
\end{figure}

As \citet{DonmezFees} note, miners receive a flat fee for Ethereum transfers between two wallets. They charge 10 to 20 times more though for smart contract interactions which include our ERC-20 tokens. Gas fees rise with network congestion, particularly after the network reaches 90\% capacity. As gas fees and Ethereum prices rise, dollar transfer costs rise with them.  I graph the trend in transfer costs for the major stablecoins in \autoref{fig:trans_fee_USD_trends}.

Between the first quarter of 2020 and the first quarter of 2022, median USD fees on Tether are up 12,550\%. USDC median fees are up 19,164\%, and Dai is up 10,899\%.  These increases reflect higher gas prices, 451\% for USDT, 770\% for USDC, and 817\% for Dai, as well as the increase in Ethereum prices.

Fees are down in the the most recent quarter, 52\% for USDC, 47\% for USDT, and 51\% for Dai.

\begin{figure}[H]
	\centering
		\caption{Transaction Cost Trends in Pct. of Transfer Size}
		\label{fig:trans_fee_pct_trends}
        \begin{minipage}{0.97\linewidth}
        \begin{center}
			\includegraphics[width=0.97\textwidth]{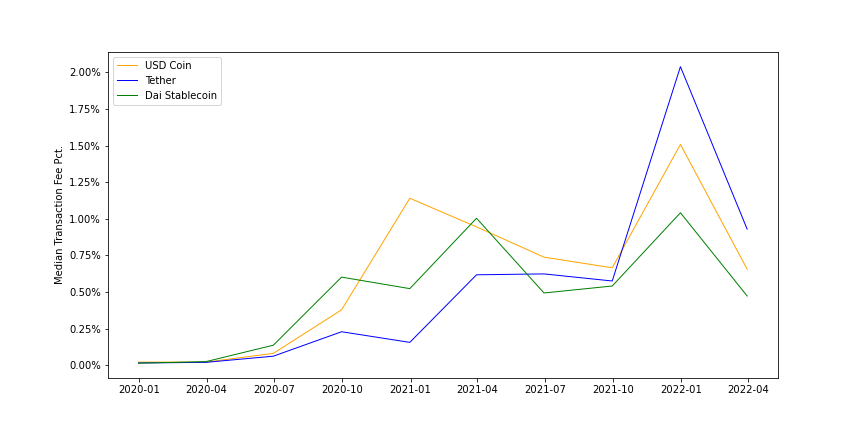} \\
		\end{center}
		\small
		\textit{Note: } I use the entire set of transactions in each quarter for each of the three major tokens and show the rise in median fees, measured in percentage of the transferred amount, over time.
    \end{minipage}
\end{figure}

The rise in median fees has made the median transfer cost in percentage terms. as shown in \autoref{fig:trans_fee_pct_trends} also much higher.  

The rise in transfer fees has already spawned blockchain innovations. 
The first is that more transactions are migrating to cheaper and faster blockchain networks.  One example is the public open source Solana.\footnote{\url{https://docs.solana.com/introduction}.  Solana reports block formation time of 695 milliseconds on May 18, 2022 at 19:34:40 GMT.  The typical transaction on Solana has fees of 5.0$\times10^{-6}$ SOL.  Even at the peak price of SOL, these fees were less than \$0.001 }  The second is the upgrade of the Ethereum network to version 2.0.\footnote{The merge of the Beacon and Ethereum chains took place on September 15, 2022. \citet{Kapengut} find, however, that the transition to proof-of-stake has not lowered transaction fees.} \citet{CataliniPOS} raise concerns about the security of new network protocols.

\section{Off-Chain Fees: Centralized Exchanges}
The primary cost to a trader off-chain is paying the exchange fees which I list in \autoref{tab:exchange_fees}.  The trader will then have to cross the bid-ask spread as well.  For stablecoins, those costs are generally much lower than the trading fee and don't impact the choice of whether to trade on or off-chain.

On-chain, a trader will pay gas fees to the miners and pool fees to liquidity providers.  I assume a 0.05\% pool fee for swaps, the middle tier fee level on Uniswap.\footnote{\url{https://docs.uniswap.org/protocol/concepts/V3-overview/fees}.}

\begin{table}[htbp]
  \centering
   \caption{Comparison of Transaction Fees 2022Q1}   \label{tab:exchange_fees} 
\begin{adjustbox}{width=1\textwidth}
    \begin{tabular}{lrrrrrrrrrrr}
\hline
          & \multicolumn{3}{c}{Off-Chain}               &       &       & \multicolumn{5}{c}{On-Chain}              &  \\
          &       &       &       &       &       & \multicolumn{2}{c}{Tether}       & \multicolumn{2}{c}{USD Coin}       & \multicolumn{2}{c}{Dai}  \\
    Volume & \multicolumn{1}{l}{Binance} & \multicolumn{1}{l}{Bitfinex} & \multicolumn{1}{l}{Coinbase} & \multicolumn{1}{l}{Huobi} & \multicolumn{1}{l}{Kraken} & \multicolumn{1}{l}{$<$min} & \multicolumn{1}{l}{$<$max} & \multicolumn{1}{l}{$<$min} & \multicolumn{1}{l}{$<$max} & \multicolumn{1}{l}{$<$min} & \multicolumn{1}{l}{$<$max} \\
\hline \hline
        Up to \$10k  & 0.10\% & 0.20\% & 0.50\% & 0.20\% & 0.26\% & \vrule{\hspace{0.08in}0.66\%} & 21.14\% & 0.42\% & 19.41\% & 1.04\% & 20.63\% \\
        \$10k - \$50k  & 0.10\% & 0.20\% & 0.35\% & 0.20\% & 0.26\% & \vrule{36.50}\% & 89.08\% & 21.93\% & 78.50\% & 23.66\% & 78.32\% \\
        \$50k - \$100k  & 0.09\% & 0.20\% & 0.25\% & 0.20\% & 0.24\% & \vrule{72.08\%} & 96.56\% & 51.47\% & 92.24\% & 38.11\% & 87.37\% \\
        \$100k - \$250k  & 0.08\% & 0.20\% & 0.20\% & 0.20\% & 0.22\% & \vrule{80.43\%} & 97.84\% & 65.80\% & 96.18\% & 45.50\% & 89.62\% \\
        \$250k - \$500k  & 0.08\% & 0.20\% & 0.20\% & 0.20\% & 0.20\% & \vrule{88.78\%} & 99.31\% & 84.07\% & 98.97\% & 66.55\% & 96.05\% \\
        \$500k-\$1m  & 0.07\% & 0.20\% & 0.20\% & 0.20\% & 0.18\% & \vrule{90.89\%} & 99.59\% & 85.05\% & 99.46\% & 71.75\% & 98.31\% \\
\hline
    \multicolumn{12}{l} {Note: For the centralized exchanges off-chain, I provide the taker fees for different volume levels.} \\
    \multicolumn{12}{l}{The table assumes on-chain swap fees of 0.05\%, the middle fee tier on Uniswap.} \\
    \multicolumn{12}{l} {I then compute the percentage of gas and swap fees in each stablecoin lower than} \\
    \multicolumn{12}{l} {the maximum and minimum off-chain fees for that activity level.} \\
    \multicolumn{12}{l} {Sources: \url{https://www.binance.us/en/fee/schedule}; \url{https://www.bitfinex.com/fees}; \url{https://pro.coinbase.com/fees};} \\
    \multicolumn{12}{l} {\url{https://www.huobi.com/support/en-us/detail}; 
\url{https://www.kraken.com/features/fee-schedule/kraken-pro}.} \\
    \end{tabular}
\end{adjustbox}
\end{table}%

The analysis by \citet{ParkMM} suggests I may be underestimating the on-chain transaction costs.  He discusses flaws in the automated market making mechanism on most DEX, and estimates they may be raising transaction costs as much as 50 basis points on a substantial fraction of trades.

\subsection{Exchange fees}
The revenue model for the five centralized exchanges is to charge both maker and taker fees to participants.  Fees are charged based on thirty-day trading volume across all trading pairs.  Fees fall with higher activity levels.  Coinbase, which charges the highest fee of 0.5\% for transactions under \$10,000, reaches a free maker tier level at \$300 million.  
Maker fees fall to zero at \$10 million on Binance, \$7.5 million on Bitfinex, and \$10 million on Kraken. On Huobi, you reach the lowest taker fee of under 0.01\% with the volume equivalent of 150,000 BTC.\footnote{Binance and Huobi offer addditional discounts for holding the exchange sponsored stablecoins.}

With the higher gas fees in the first quarter of 2022, Mainnet transactions are generally uncompetitive for retail size transactions.  For Tether, only 0.66\% of transfers have costs below the 0.1\% on Binance, and only 21.1\% are lower than the much higher 0.5\% on Coinbase.  The numbers are similar, but slightly lower, for USD Coin and Dai.  Less than 21\% of transfers in those two coins are lower than the 0.5\% on Coinbase.

By the time you reach the third activity level, \$50,000 to \$100,000, the majority of transactions in Tether and USD Coin are lower than the 0.09\% fee on Binance.  By the time you reach the \$250,000 to \$500,000 level the majority of on-chain gas and pool fees are lower than the exchange fees.  For Tether and USD Coin, it is more than 84\%.

\subsection{Bid-ask spreads}
The second component of off-chain exchange costs is the bid-ask spread.\footnote{I abstract from the execution quality which could potentially be worse off-chain.}  Those costs are generally neglible compared to the exchange fees.

I study a variety of pairs for each of the three major stablecoins for the last year April 1, 2021 to March 31, 2022. The four pairs I study for Tether are BUSD/USDT on Binance, USDT/USD on Bitfinex, USDT/HUSD on Huobi, and USDT/USD on Kraken.  For USD Coin, I analyze three pairs: USDC/USDT on Binance, BTC/USDC on Coinbase, and USDC/HUSD on Huobi.  I analyze four pairs of Dai, DAI/USD on Bitfinex, Coinbase, and Kraken and DAI/USDT on Huobi.

\begin{figure*}
\caption{Median Bid-Ask Spreads (Pct.) by Exchange of Major Stablecoins}
\label{fig:stablecoin_bid_ask_spreads}
\begin{multicols}{2}
\center (a) Tether
\includegraphics[width=\linewidth]{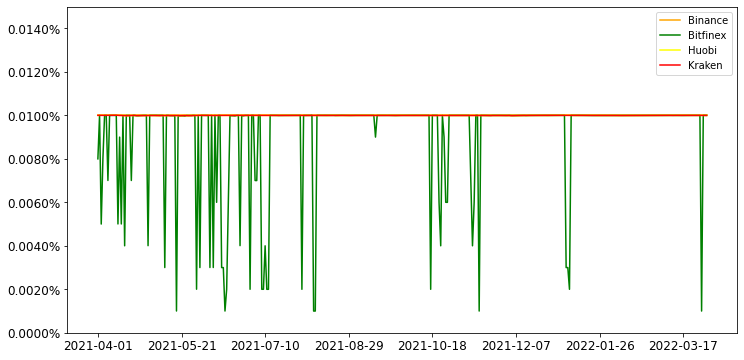}
\center (b) USD Coin
\includegraphics[width=\linewidth]{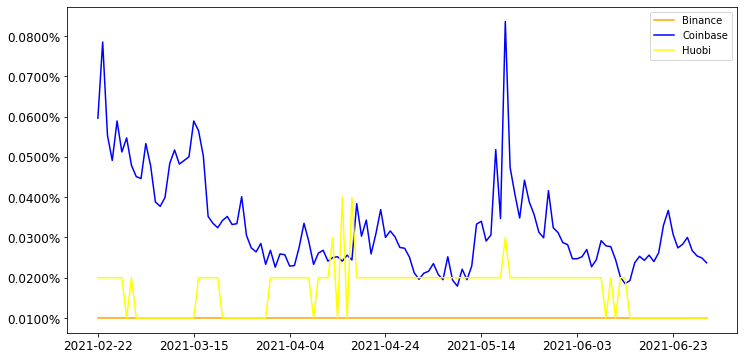}\par
\end{multicols}
\center (c) Dai \\
\includegraphics[width=0.5\linewidth]{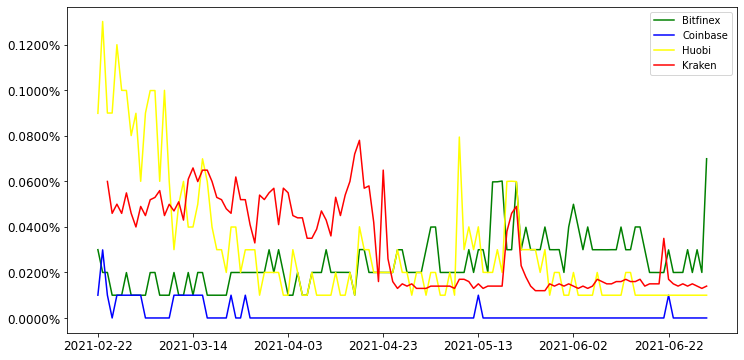}\par 
\textit{Note:} The best-bid or offer is computed from direct API feeds for each exchange. 
\end{figure*}

I next report the median bid-ask spreads for the four currencies by exchange-pair in  \autoref{fig:stablecoin_bid_ask_spreads}.

All of the exchanges maintain generally stable bid-ask spreads of 0.01\% or lower in Tether.  Bitfinex is most often the tightest.  With Tether trading near \$1.00, this is a cost of 0.0001\$ for most transactions.

I consider USD Coin in two categories.  On Binance, you exchange USD Coin for Tether.  On Huobi, you are exchanging for Houbi USD.  On Coinbase, which sponsors USD Coin and provides unlimited parity trades\footnote{\url{https://www.coinbase.com/usdc}: ``For customers with a US dollar bank account, one USDC can always be redeemed for US\$1.00.} between the coin and the USD, you are exchanging for Bitcoin.  Binance bid-ask spreads are uniformly at the typical 0.01\%.  Huobi is frequently double that at 0.02\%.  On Coinbase, since you are trading for a risky asset, the bid-ask spread of 0.027\% is similar to the BTC-USD rate.

Bid-ask spreads on Dai are similar to Tether on Coinbase, 0.005\% and Kraken, 0.011\%.  Btifinex and Houbi are twice as wide as Kraken though.

The median bid-ask spreads have little impact on comparative transaction costs in \autoref{tab:exchange_fees}.  I add 0.01\% to the maximum fees for all three stablecoins.  For Tether, the on-chain preference changes by less than 0.5\% in each category.  For USD Coin it rises by 0.4\% on average, with the largest increase of 0.82\% in the \$10,000 to \$50,000 range.  Dai is preferred 0.6\% more on average, with a 1.1\% increase in the 100,000-250,000 category.

\begin{figure*}
\caption{95th Percentile Bid-Ask Spreads by Exchange of Major Stablecoins}
\label{fig:stablecoin_bid_ask_p95spreads}
\begin{multicols}{2}
\center (a) Tether
\includegraphics[width=\linewidth]{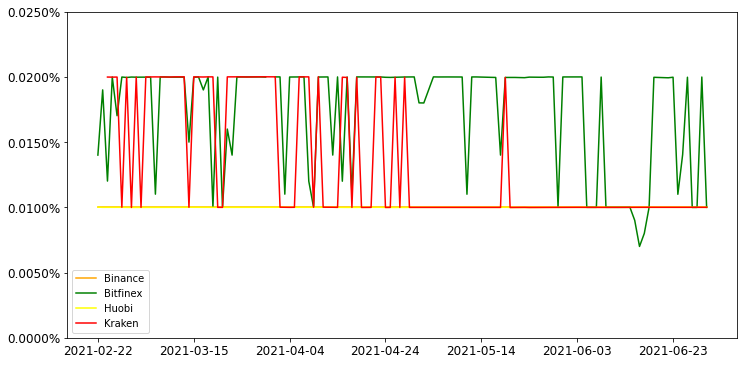}
\center (b) USD Coin
\includegraphics[width=\linewidth]{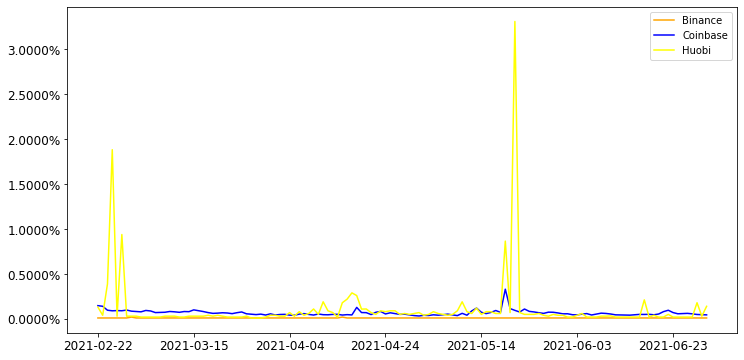}\par
\end{multicols}
\center (c) Dai \\
\includegraphics[width=0.5\linewidth]{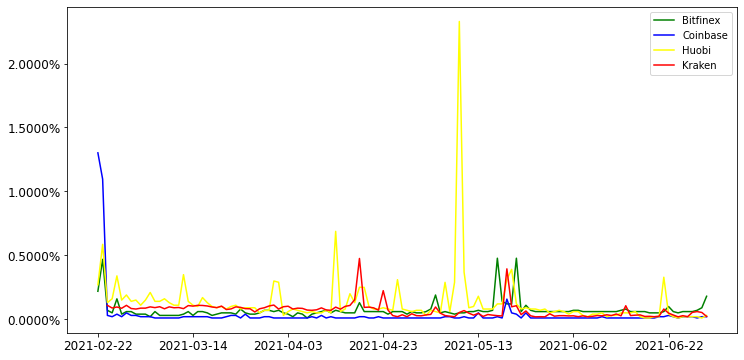}\par 
\textit{Note:} The best-bid or offer computed from direct API feeds from each exchange. 
\end{figure*}

On occasion, the bid-ask spreads increase significantly though.  I plot in \autoref{fig:stablecoin_bid_ask_p95spreads} the 95th percentile of bid-ask spreads for each of the currency-exchange groups. 

With Tether, even trading on the exchange with the widest spreads, Kraken, a trader would face a spread of 0.02\% or more less than 5\% of the time.  This is quite important because as \citet{BianchiTether} show, deviations in Tether from parity impact the out-of-sample returns on a broad index of digital currencies.

Huobi has occasional break downs in their usual tight spreads for USD Coin.  5\% of the time you would trade against a bid-ask spread of more than 3.31\%.

With Dai, on Bitfinex, bid-ask spreads occasionally exceed 0.5\%.  On Huobi, they sometimes exceed 2.0\%.  On Coinbase, the 95th percentile spreads never exceed 0.2\%.

I compare these less frequent outcomes to the tail distributions for gas fees.  In the smallest transaction category, 5\% of Tether transactions consume 44.8\% of the transfer amount.  The corresponding numbers are 61.6\% for USD Coin and 93.9\% for Dai.  What makes these numbers worse is that you can pay gas fees even for unsuccessful transactions.

\subsection{Depth}

While spreads effect transactions costs, they might be a poor estimate if the markets are not very deep.

\begin{figure*}
\caption{Median Depth (USD thsd.) by Exchange of Major Stablecoins}
\label{fig:stablecoin_depth}
\begin{multicols}{2}
\center (a) Tether
\includegraphics[width=\linewidth]{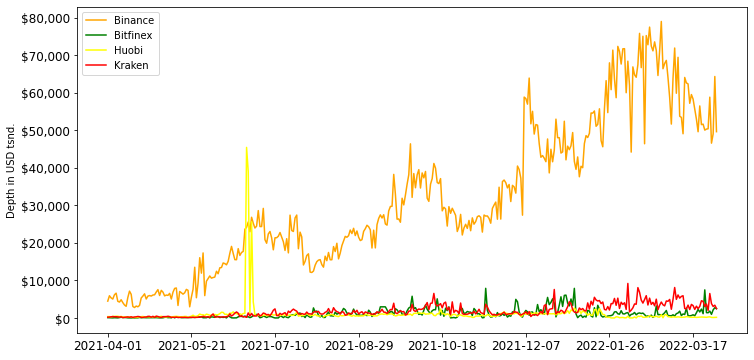}
\center (b) USD Coin
\includegraphics[width=\linewidth]{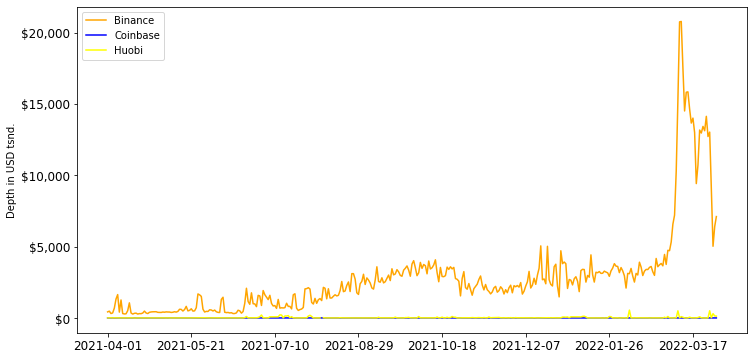}\par
\end{multicols}
\center (c) Dai \\
\includegraphics[width=0.5\linewidth]{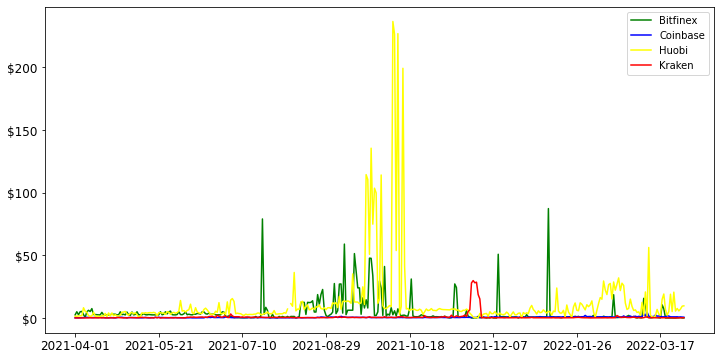}\par 
\flushleft
\textit{Note:} The depth (in thousands of USD) is one-half the sum of displayed depth at the best-bid and offer computed from direct API feeds for each exchange. 
\end{figure*}

I calculate depth as
\begin{center}
0.5 $\times$ [bid depth + ask depth],
\end{center}
in thousands of USD for the five currency-exchange groups.  \autoref{fig:stablecoin_depth} provides data on the token-exchange pairs.

Tether is by far the deepest market.  Binance displays a median depth that averages more than 31 million Tether. Depth was increasing steadily in 2021 reaching nearly \$70 million. Huobi is the second deepest in Tether, but it is not even 5\% as deep as Binance.  The only coin with comparable exchange depth is USD Coin on Binance, which has a median depth which averages 2.84 million.

Dai, on Bitfinex, Coinbase, and Kraken, would see market orders of more than 5,000 Dai likely breaking through the best bid or offer; the largest depth is 11.6 thousand on Huobi.

\section{Speed}

The Ethereum Mainnet is too slow for any type of low-latency trading on DEX.  In centralized exchanges, I find high frequency trading activity that is similar to the equity market.

\subsection{On-chain}
\begin{figure}[H]
	\centering
		\caption{Price and Speed in the MetaMask Wallet App}
		\label{fig:wallet_app}
        \begin{minipage}{0.47\linewidth}
        \begin{center}
			\includegraphics[width=0.47\textwidth]{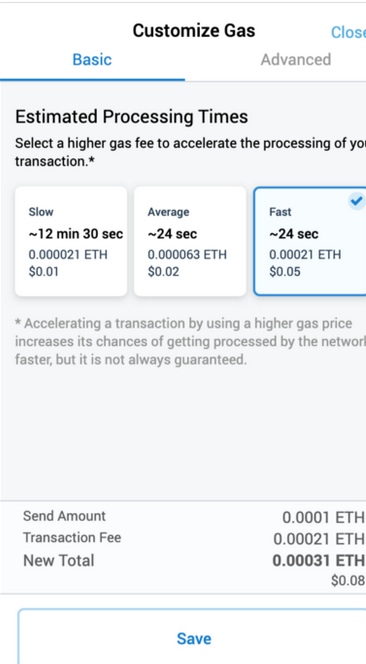} \\
		\end{center}
		\small
    \end{minipage}
\end{figure}

The speed in which any transfer gets executed is a function of the price you are willing to pay.  In \autoref{fig:wallet_app}, I show a screen shot from the popular Metamask\footnote{\url{https://metamask.io/}} Wallet App.

\textit{Note: } The example is from \citet{Winfrey} who provides a schematic of the Metamask algorithm.
\vspace{0.25in}\\
The app estimates network congestion and provides expected times for the transaction to complete based on the amount you are willing to pay. The risk with setting the price too low is that your transaction may never complete.

\citet{WernerGas} fit a deep-learning model to more than 500,000 blocks on the Mainnet.  Compared to the Geth\footnote{Based on statistics from \url{https://www.ethernodes.org/}, retrieved on February 5, 2022, 82\% of Mainnet users utilize the Geth client} client, their model achieves a 50\% savings on gas, while only slowing down the transaction by 1.3 blocks on average.  While this may seem inconsequential, this is an average delay of nearly twenty seconds, an eternity in the low latency trading environment.

Etherscan provides\footnote{\url{https://etherscan.io/chart/blocktime}} a time series of the average transfer speed in the block chain. The median over 2015-2021 is 14.13 seconds, with a maximum of 30.31 seconds.  This data is helpful, but it does not address the fact that some transactions can take days or weeks or sometimes never complete.  There is also no guaranteed time delivery, regardless of rate.

\subsection{High frequency trading on centralized exchanges}

High-frequency trading (HFT) has become an important part of the digital asset trading environment.  \citet{MizrachBTC} found HFT market share at Coinbase of 15.5\% for Bitcoin.  The authors did not examine stablecoins though, and they looked at a different set of exchanges apart from Coinbase.  Recent industry discussion suggests that HFT activity has increased, with the arrival of equity HFT firms like Jump Trading and DRW.  Industry estimates suggest that 80 to 90\% of the trading on Bitfinex is from HFT firms.\footnote{\url{https://www.supercryptonews.com/over-80-of-trading-volume-on-bitfinex-is-generated-by-hft-firms/}:“From Jump to DRW, a lot of HFT firms are diving in headfirst because of a lack of opportunities in the traditional high frequency, low latency trading.”}

\begin{figure*}
\caption{Cancellation/Execution Ratios of Major Stablecoins}
\label{fig:stablecoin_ceratio}
\begin{multicols}{2}
\center (a) Tether
\includegraphics[width=\linewidth]{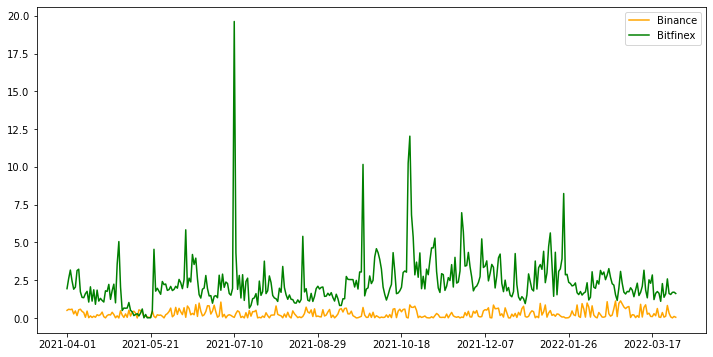}
\center (b) USD Coin
\includegraphics[width=\linewidth]{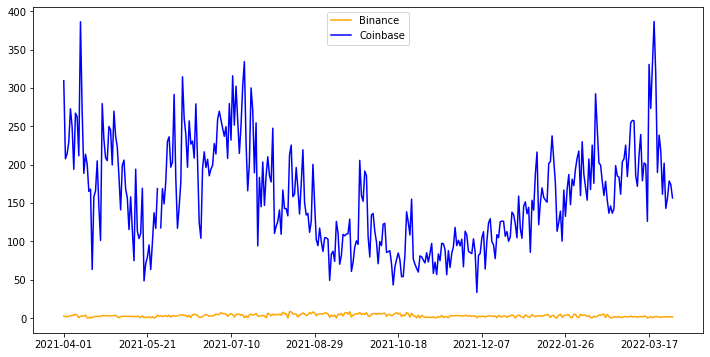}\par
\end{multicols}
\center (c) Dai \\
\includegraphics[width=0.5\linewidth]{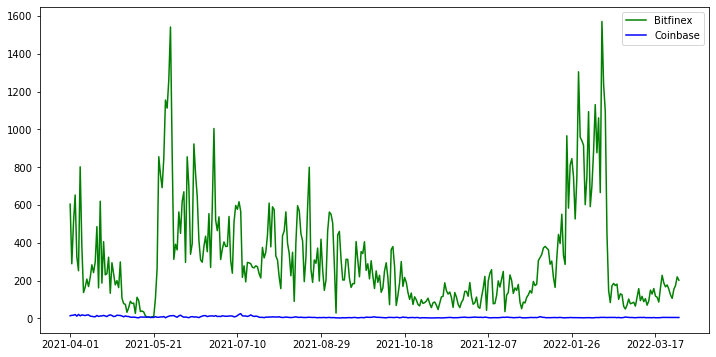} \\
\textit{Note:} Cancellations are computed from direct API feeds from each exchange.  
\end{figure*}

One measure of HFT activity is the cancellation to execution ratio.  I can compute these estimates for only a subset of the exchanges, Binance, Bitfinex and Coinbase, where the API provides accurate timestamps of the orders.  These estimates are in \autoref{fig:stablecoin_ceratio}.

Cancellation to execution (CE) ratios are under three on Binance for Tether and USD Coin.  This is low compared to the typical US large cap equity.  The SEC Midas website\footnote{\url{https://www.sec.gov/marketstructure/datavis/ma_stocks_canceltotrade.html}} reports a 22-day moving average of 9.52 for January 2022.  Along with the narrow spreads, this suggests liquidity providers face little risk of adverse selection.

Dai has higher CE ratios: over six on Coinbase and over 300 on Bitfinex.

\begin{figure*}
\caption{HFT Pct. of Order Cancellations}
\label{fig:stablecoin_hftcancel}
\begin{multicols}{2}
\center (a) Tether
\includegraphics[width=\linewidth]{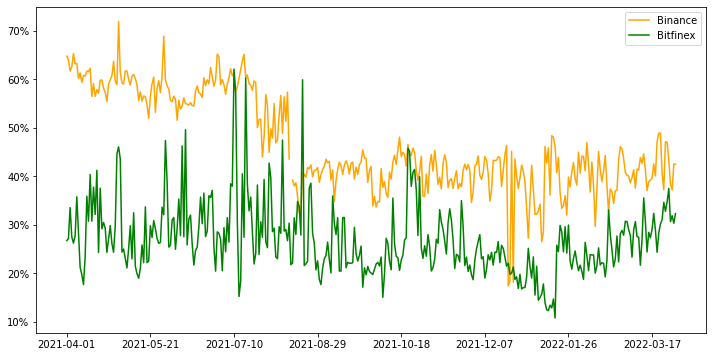}
\center (b) USD Coin
\includegraphics[width=\linewidth]{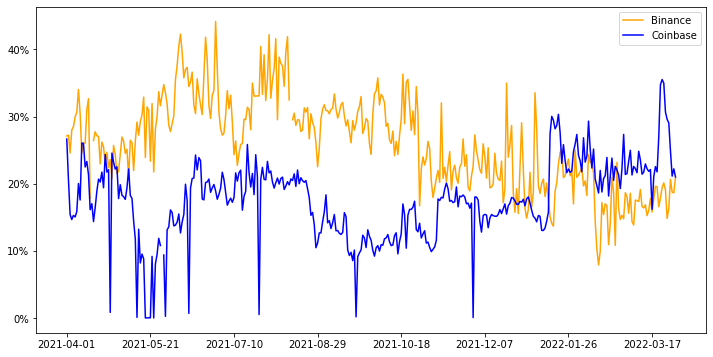}\par
\end{multicols}
\center (c) Dai \\
\includegraphics[width=0.5\linewidth]{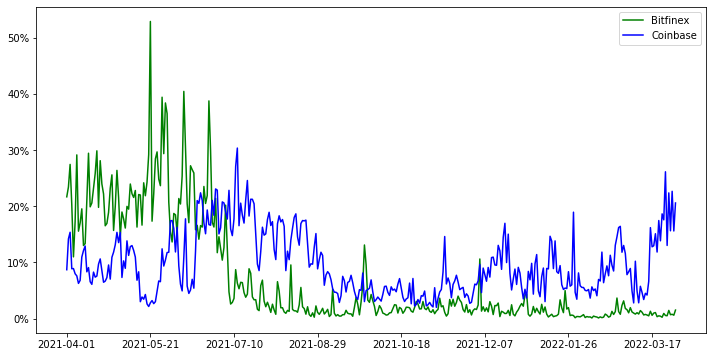}\par 
\textit{Note:} Cancellations are computed from direct API feeds from each exchange. Thresholds are 500 milliseconds on Binance and Bitfinex and 50 millseconds on Coinbase. 
\end{figure*}

A second measure is the percentage of cancellations under 500 milliseconds.  These are orders which are lifted (either through trades or cancellation) from the order book before any non-HFT trader could react to them. I report estimates in \autoref{fig:stablecoin_hftcancel}.

My highest share of HFT activity is 46.7\% for Tether on Binance.  Tether on Bitfinex and USD Coin on Binance are over 25\%.  Coinbase has less HFT activity.  17.5\% of USD Coin cancellations are below the HFT threshold.

The HFT shares for Dai are lower, 9.7\% on Coinbase and 7.2\% on Bitfinex.

There appear to be a variety of motivations for the HFT trading in stablecoins.  The first is to arbitrage small price discrepancies on an exchange or across exchanges.\footnote{Ryan Adams, ``How to make money trading stablecoins,'' \url{lhttps://newsletter.banklesshq.com/p/how-to-make-money-trading-stablecoins}}  The second is as an on-exchange store of value.  Getting assets into an exchange to exploit trading opportunities can be quite slow.  Coinbase, for example, requires 35 confirmations for a token transfer, so it can easily take ten minutes to transfer stablecoins between exchanges.\footnote{\url{https://suredbits.com/lightning-201-high-frequency-trading/}}

There is an open debate on whether the HFT activity is stabilizing for the digital asset markets, see e.g. \citet{Alexander}, but it is clear that stablecoins will be a part of that discussion.

\section{Conclusion}

I have analyzed the collateral structure of the ten leading stablecoins by market capitalization.  Seven of the ten coins are backed primarily by cash equivalents assets.  Three hold digital assets, and two of those are algorithmically stabilized.  Intra-daily price variation is smaller for the well-capitalized stablecoins like Tether and USD Coin that are backed by cash equivalents.

While this paper has analyzed the successful tokens, failure rates of stablecoin projects are almost as high for other digital assets. Since 2016, more than 60\% of stablecoins have failed.  The survivorship rate is similar to other non-stable tokens on the Mainnet.

I estimate  \$1.47 trillion in stablecoin transactions in the first quarter of 2022. USD Coin leads Ethereum Mainnet stablecoin transactions, with a 48\% market share.  Tether is second at 32\%.    Dai has the highest velocity at 16.8.  USD Coin is second at 15.4, and Tether is third at 11.8.  These are two to three times the M1 velocity because most transfers are exchanges for other digital assets.

The Ethereum blockchain provides a great deal of transparency at the level of the hash tag address.  In the first quarter of 2022, Tether transactions involve 2.2 million unique counterparties. USD Coin has 1.16 million. Dai has 177,000. The remaining tokens have fewer than 50,000 counterparties.  Decentralized and centralized exchanges are by far the most active network addresses on the blockchain.  I compute Herfindahl indices for the stablecoin network. Binance USD and Gemini have the most concentrated holdings.  For both, the largest holder has more than 50\% of the supply.  Along with Huobi USD, these three tokens would be considered ``highly concentrated" under DOJ merger guidelines.

The total fees from 2022Q1 for the top ten stablecoins are are \$336.80 million. USD Coin
generates \$148 million in fees (44.0\%), Tether \$158 million (43.9\%), and Dai, \$30 million
(8.9\%).  The median transaction fee for Tether is \$13.03 or 0.93\% of the transferred amount. 1.78\% of transfers have fees which exceed the transferred amount.  USD Coin and Dai have higher median transfer fees, \$18.84 for USDC and \$19.80 for Dai.  For both of these stablecoins, more than 25\% of transactions are in excess of 3\% of the amount transferred.  To compete with existing payment systems on fees, retail size transfers must transact off the Mainnet.

A major obstacle for the Mainnet is that network congestion is raising gas fees.  As the price of Ethereum has risen, so have transfer fees.  Median fees for Tether rose over 12,500\% over the last two years, They rose over 19,000\% for USD Coin, and almost 11,0000\% for Dai.

On September 15, 2022, the Ethereum Mainnet completed its transition from proof-of-work to proof-of-stake. While this has dramatically lowered the energy consumption of the network, \citet{Kapengut} have found that transfer fees actually rose more than 20\% for tokens. 

Transactions of under \$10,000 are cheaper on centralized exchanges more than 95\% of the time.  Gas fees do not rise proportionally with transfer sizes, so for transactions of more than \$50,000, the Mainnet is less expensive the majority of the time.  Blockchains with lower transaction costs like Solana have become important competitors to the Mainnet on both fees and speed.

The last stage of my analysis shows that the major digital asset exchanges provide low cost, deep markets in the major stablecoins.  24 hour turnover in Tether is nearly \$60 billion.  This is more than double the daily turnover of all the FANG stocks in March 2022 (Facebook now Meta, Amazon, Netflix, and Google, now Alphabet).  It is eight times the daily flow in money market mutual funds.\footnote{The average outflow from the sector was \$484.4 billion in 2021Q4, https://fred.stlouisfed.org/series/ MMMFTAQ027S.  This averages to \$7.34 billion per day, assuming 66 trading days.}

The \citet{Basel} has started the discussion on capital requirements for digital assets.  The seven tokens backed by cash equivalents may qualify as Group 1a assets requiring little or no reserves.  The three tokens that hold digital assets would likely fall into Group 1b with much higher capital set asides.

The President's Working Group on Financial Markets\footnote{\citet{WorkingGroup}} has proposed that stable coin issuers become insured depository institutions.  There is an ongoing debate, see e.g. \citet{Malloy}, about whether central banks should offer competing digital currencies and how they should be funded.  
The instability of the private offerings should impact both discussions.

\pagebreak
\singlespace


\begingroup
\setstretch{1.25}
\printbibliography
\endgroup

\end{document}